%% file: main.tex
\documentclass[letterpaper,twocolumn,10pt]{article}

\usepackage{tikz}
\usepackage{amsmath}
\usepackage[table]{xcolor}

\usepackage{filecontents}
\usepackage{preamble}

\usepackage{booktabs}
\usepackage{tabularx}

\input{macros}

\newif\ifcomments
 \commentstrue 

\ifcomments
    \providecommand{\tgriggs}[1]{{\color{cyan}{tgriggs: #1}}}
    \providecommand{\accheng}[1]{{\color{purple}{accheng: #1}}}
    \providecommand{\soujanya}[1]{{\color{olive}{soujanya: #1}}}
    \providecommand{\jw}[1]{{\color{yellow}{jae: #1}}}
    \providecommand{\matei}[1]{{\color{teal}{matei: #1}}}
    \providecommand{\wenjie}[1]{{\color{magenta}{\textbf{wenjie:} #1}}}
    \providecommand{\devbali}[1]{{\color{red}{devbali: #1}}}
    \providecommand{\ion}[1]{{\color{magenta}{ion: #1}}}
\else
    \providecommand{\tgriggs}[1]{}
    \providecommand{\accheng}[1]{}
    \providecommand{\soujanya}[1]{}
    \providecommand{\jw}[1]{}
    \providecommand{\matei}[1]{}
    \providecommand{\wenjie}[1]{}
    \providecommand{\devbali}[1]{}
    \providecommand{\ion}[1]{}
\fi

\begin{document}

\makeatletter
\def\blfootnote{\xdef\@thefnmark{}\@footnotetext}
\makeatother

\date{}


\title{\newRFfull: Performance Isolation for Multi-Tenant Storage Systems}
\author{
{\rm Tyler Griggs\textsuperscript{*}, Soujanya Ponnapalli\textsuperscript{*}, Dev Bali, Wenjie Ma, James DeLoye, Audrey Cheng, Jaewan Hong,}
\and 
{\rm Natacha Crooks, Scott Shenker, Ion Stoica, Matei Zaharia}
\\University of California, Berkeley
}
\maketitle
\blfootnote{*These authors contributed equally}

\input{00-abstract}

\pagestyle{plain}

\input{01-introduction-osdi26}
\input{02-background-osdi26}

\input{03-fairness-osdi26}

\input{04-fairdb}
\input{06-impl}

\input{07-eval}

\input{08-related}

\input{09-conclusion}
\bibliographystyle{plainurl}
\bibliography{references}

\clearpage


\end{document}

%% file: macros.tex
\newcommand{\io}{\textsc{I/O}\xspace}
\newcommand{\ie}{\emph{i.e.,}\xspace}

\newcommand{\newRFfull}{Delta Fair Sharing\xspace}
\newcommand{\newRFsymbol}{\ensuremath{\delta}\xspace}

\newcommand{\resBuffer}{\ensuremath{\rho_{\mathrm{buffer}}}}
\newcommand{\resCache}{\ensuremath{\rho_{\mathrm{cache}}}}

\newcommand{\worstcasedemand}{worst-case demand\xspace}
\newcommand{\peakdemandthreshold}{ramp-up threshold\xspace}

\newcommand{\sys}{\textsc{FairDB}\xspace}

\newcommand{\propfull}{Delta Fair Sharing\xspace}

%% file: 00-abstract.tex
\begin{abstract}
Modern storage systems, often deployed to support multiple tenants in the cloud, must provide performance isolation. Unfortunately, traditional approaches such as fair sharing do not provide performance isolation for storage systems, because their resources (e.g., write buffers and read caches) exhibit high preemption delays. These delays lead to unacceptable spikes in client tail latencies, as clients may be forced to wait arbitrarily long to receive their fair share of resources.

We introduce \newRFfull, a family of algorithms for sharing resources with high preemption delays. These algorithms satisfy two key properties: \newRFsymbol-fairness, which bounds a client’s delay in receiving its fair share of resources to \newRFsymbol time units, and \newRFsymbol-Pareto-efficiency, which allocates unused resources to clients with unmet demand. Together, these properties capture resource-acquisition delays end-to-end, bound well-behaved clients' tail-latency spikes to \newRFsymbol time units, and ensure high utilization. We implement such algorithms in \sys, an extension of RocksDB. Our evaluation shows that \sys isolates well-behaved clients from high-demand workloads better than state-of-the-art alternatives.

\end{abstract}

%% file: 01-introduction-osdi26.tex
\section{Introduction}
\label{sec:intro}

Multi-tenancy enables cloud service providers to improve resource utilization and reduce costs. To achieve these goals, multi-tenant storage systems~\cite{rocksdb, lakshman2010cassandra, mysql, postgresql} share physical cloud resources and support multiple, diverse clients across tenants~\cite{cheng2018analyzing, reiss2012heterogeneity, yang2021large}. In such storage systems, ensuring \textit{performance isolation}—where one client’s behavior does not impact the performance of others—remains a fundamental challenge. The lack of isolation can lead to severe performance degradation: a single client with a high-demand workload can cause up to 977$\times$ worse tail latency for well-behaved clients across systems such as PostgreSQL~\cite{postgresql} and RocksDB~\cite{rocksdb}, and even the underlying Linux OS page cache, as shown in \autoref{tab:intro-mot} (\S\ref{sec:bg-and-mot}).

\begin{table}[t]
\centering
\small
\begin{tabular}{lcc}
\hline
\multirow{2}{*}{\textbf{\begin{tabular}[c]{@{}l@{}}System\\ Allocation Policy\end{tabular}}} &
  \multirow{2}{*}{\textbf{\begin{tabular}[c]{@{}c@{}}With no\\ interference\end{tabular}}} &
  \multirow{2}{*}{\textbf{\begin{tabular}[c]{@{}c@{}}With one high-\\ demand client\end{tabular}}} \\
                       &                       &        \\ \hline
OS Page Cache            & 3.28 ms  & 45.82 ms \hfill (14$\times$)\\
PostgreSQL                                       &  5.3 ms     &  323 ms \hfill (61$\times$)\\
RocksDB                  &    \textasciitilde1ms      &  977 ms \hfill (977$\times$)\\
RocksDB (Fair sharing)   &     \textasciitilde1ms     &   246ms \hfill (246$\times$)\\ \hline
\end{tabular}
\caption{Impact of a single high-demand client (relative to none) on the p99 latency of well-behaved clients across storage systems. The experiment setup and workloads are outlined in \S\ref{sec:bg-and-mot}.}
\label{tab:intro-mot}
\end{table}

To achieve performance isolation, fair sharing has been widely applied in other domains, ranging from CPU scheduling and cluster scheduling to networking~\cite{demers1989analysis,drf, drfq, vuppalapati2023karma, waldspurger1994lottery, waldspurger1995stride, linux_cfs}. This approach ensures that each client can receive a fair share of resources (for performance isolation), while unused shares are redistributed to clients with high demand (improving utilization). However, fair sharing makes a fundamental assumption: resources can be quickly redistributed among clients, i.e., resources with \textit{negligible preemption delays}. For instance, in networking, link bandwidth can be redistributed after the current packet is transmitted (within a few $\mu$s), allowing clients to rapidly converge to their fair shares.

Unfortunately, fair sharing does not guarantee performance isolation for storage systems. Redistributing resources in datastores, be it as write buffers and read caches, requires multiple slow disk \io operations—i.e., storage resources have \textit{high preemption delays}. As a result, clients may face unbounded delays to receive their fair share of resources and have unacceptable spikes in their tail latencies. We see this in practice: despite enforcing fair sharing of resources in RocksDB, a single client with a high-demand workload can cause up to 246$\times$ worse tail latency for well-behaved clients, as shown in \autoref{tab:intro-mot} (\S\ref{sec:bg-and-mot}). Clients reclaiming their share of the write buffer must wait for other high-demand clients to free up buffer space by flushing their dirty pages to disk, which can take hundreds of milliseconds. Likewise, a client reclaiming its share of an in-memory read cache must wait for expensive disk reads to repopulate its cache (i.e., fetch data that has been evicted by other clients).
\changebars{}{\accheng{stateful storage resources?}} 

\begin{figure}[t]
    \includegraphics[width=0.99\linewidth]{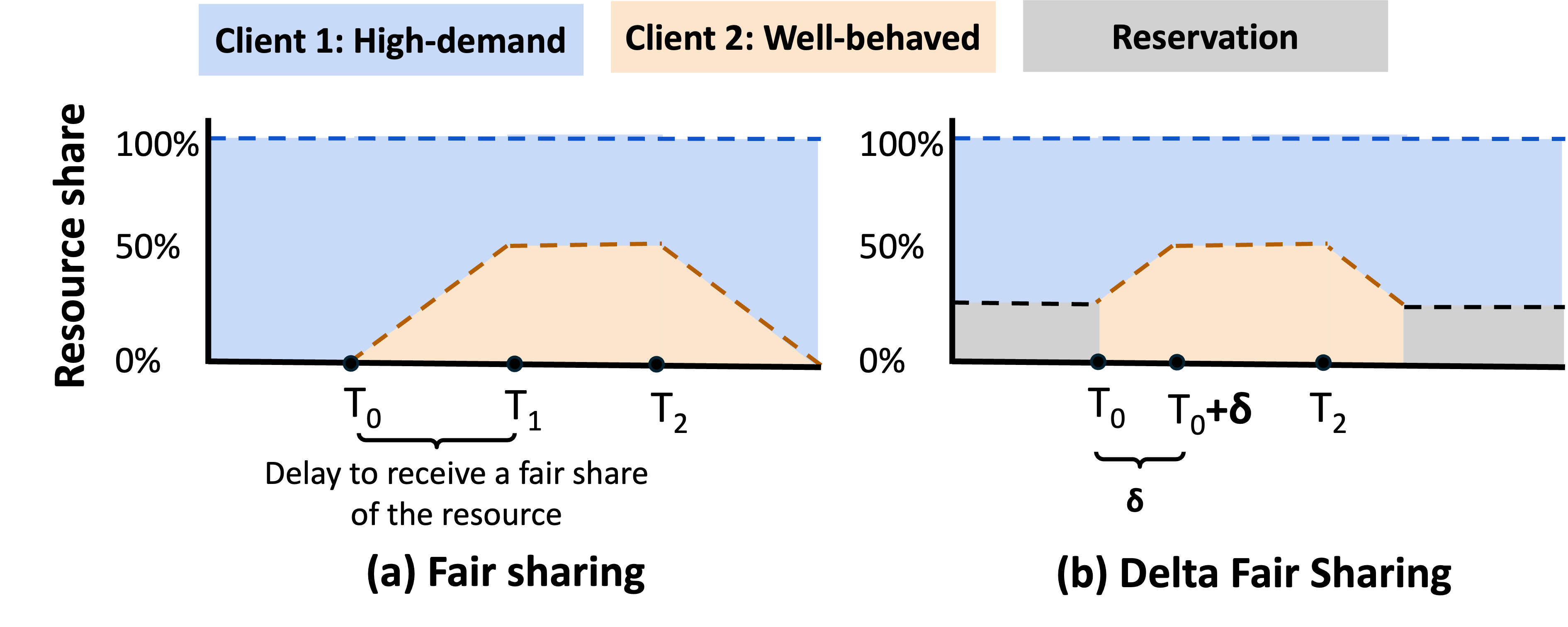}
    \caption{Example. A well-behaved client demands its share at T$_0$ and experiences long delays to receive its share with traditional fair sharing. \newRFfull bounds such delays (to \newRFsymbol time units).}
    \label{fig:example}
\end{figure}

\changebars{}{
\accheng{There is a disconnect between the first half of the intro and the sections below. The first half focuses on challenges due to high preemption delays -- we need to tie this explicitly to the two key properties, resource reservations, and \newRFfull being introduced. I've included a possible restructuring below, feel free to edit.\soujanya{Few considerations that went into the current intro structure: the definition of \newRFfull and its properties must be general, reservations are a mechanism of \textit{our} \newRFfull algorithms, and its structure is consistent with the rest of the paper.}

The root cause of this interference is that standard fair sharing is \textit{work-conserving}: it eagerly redistributes all unused resources to maximize utilization under the assumption that they can be recovered with negligible preemption delays. However, this assumption fails when applied to storage systems, so the eager redistribution of resources breaks performance isolation. Resources are lent out but cannot be reclaimed in time, causing well-behaved clients to be delayed. Consequently, storage systems face a trade-off between performance isolation and high utilization.

To resolve this, we make the key insight that the system must strategically withhold, or \textit{reserve}, a portion of resources to ensure they remain immediately available for well-behaved clients. Based on this, we introduce \newRFfull, a family of resource-allocation algorithms designed to guarantee performance isolation even with have high preemption delays. \newRFfull manages the trade-off between isolation and utilization through two key properties: \newRFsymbol-fairness and \newRFsymbol-Pareto-efficiency. First, \newRFsymbol-fairness ensures performance isolation by bounding the time a client must wait to receive their fair share to at most \newRFsymbol time units. This \newRFsymbol parameter is configurable, allowing system administrators to define a maximum delay tolerance similar to rate limits in commercial storage systems~\cite{aws-rds,bigtable_quotas} (\S\ref{sec:rad_definition}). 
Second, \newRFsymbol-Pareto-efficiency ensures high utilization. While isolated systems with fixed per-client resource quotas trivially satisfy \newRFsymbol-fairness, they incur poor utilization. Thus, this property allows the system to redistribute unused resources so long as the \newRFsymbol-fairness bound is preserved.
This formulation generalizes existing approaches: when \newRFsymbol is set to zero, \newRFfull reduces to isolated systems with fixed resource quotas, and as \newRFsymbol approaches $\inf$, it converges to traditional fair sharing.

To satisfy these properties, \newRFfull relies on \textit{dynamic resource reservations}. If a resource takes a long time to preempt (e.g., $>$\newRFsymbol), the scheduler must maintain a ``headroom'' of free capacity. 
Specifically, a base amount of that resource must be withheld from redistribution to other clients, i.e., reserved, even if unused, so that a client can reclaim the remaining portion and receive its fair share within \newRFsymbol time units, as shown in \autoref{fig:example}. 
However, determining the necessary reservation size to satisfy \newRFsymbol-fairness is difficult because preemption latency is both resource- and workload-dependent. For example, the amount of write buffer a client can reclaim within \newRFsymbol depends on the available disk bandwidth, which in turn depends on the number of active clients and other system events, such as compactions or garbage collection, that demand bandwidth. To address this, we design algorithms that compute \textit{minimal reservations} to account for system events with respect to a \textit{\worstcasedemand} model based on historical patterns~\cite{snowflake-nsdi20, yang2021large}.
}
}

In this paper, we introduce \newRFfull, a family of resource-allocation algorithms that address the problem of sharing resources with high preemption delays to provide performance isolation for storage systems. These algorithms satisfy two key properties: \newRFsymbol-fairness and \newRFsymbol-Pareto-efficiency. The \newRFsymbol-fairness property introduces a maximum delay~(\newRFsymbol) and ensures that clients receive their fair share of resources within \newRFsymbol time units. In practice, system administrators can configure this maximum delay bound, similar to SLOs in commercial storage systems~\cite{aws-rds,bigtable_quotas}\footnote{When \newRFsymbol is set to zero, \newRFfull reduces to an isolated system with fixed per-client resource quotas, and when \newRFsymbol is unbounded ($\newRFsymbol=\inf$), \newRFfull encompasses traditional fair-sharing algorithms.}. The \newRFsymbol-fairness property ensures that a client’s delay to receive its fair share does not exceed \newRFsymbol time units, even under \textit{\worstcasedemand} conditions.

The \newRFsymbol-fairness property alone could trivially be satisfied by reserving a static amount of resources for each client and keeping it idle in case they are in demand, which does not capture the other benefit of fair sharing, that clients can consume unutilized resources from others.
To address this, the \newRFsymbol-Pareto-efficiency property requires redistributing unutilized resources to clients with unmet demand, as long as the system can still provide \newRFsymbol-fairness under some \worstcasedemand conditions.
In the extreme, if we consider the ``worst case'' as all clients simultaneously requesting their full fair share of resources, then this property would still require keeping most resources idle and lead to low utilization.
However, \newRFsymbol-Pareto-efficiency can be defined with other conditions.
In this paper, we consider a simple \worstcasedemand condition that captures the maximum number of clients that can concurrently ramp up to their fair share—i.e., a \textit{\peakdemandthreshold} ($k$). In most multitenant workloads, most clients are not active all the time, so $k$ can be set low.
Moreover, $k$ can easily be set based on historical patterns~\cite{snowflake-nsdi20, yang2021large} by monitoring how many clients have ever ramped up simultaneously in the past.
Our algorithms for \newRFfull enable systems to achieve both \newRFsymbol-fairness and high utilization as long as the assumed \worstcasedemand conditions are met. \newRFfull naturally extends to other \worstcasedemand conditions.

We design such \newRFfull algorithms for two key storage resources—write buffers and read caches—which are the primary sources of tail-latency spikes in RocksDB. Our algorithms withhold a portion of a client’s fair share from redistribution to other clients—i.e., make \textit{reservations}—so that the client can reclaim the remaining amount within \newRFsymbol time units even under \worstcasedemand, as illustrated in \autoref{fig:example}. Each algorithm computes the minimum reservation size needed to satisfy \newRFsymbol-fairness (with the maximum delay \newRFsymbol) and achieves \newRFsymbol-Pareto-efficiency assuming the \peakdemandthreshold ($k$) based on that resource's implementation and reallocation dynamics.


We implement our algorithms in \sys, a prototype that extends the popular RocksDB key-value store with \newRFfull (\S\ref{sec:policies-rocksdb}). We showcase its benefits in a standard multi-tenant deployment with independent client workloads. Our evaluations show that \sys bounds the end-to-end tail-latency spikes of well-behaved clients to \newRFsymbol time units and isolates them from extremely high-demand workloads while achieving high utilization. On production traces~\cite{snowflake-nsdi20,ycsbpaper}, \sys achieves up to \textbf{9$\times$} better tail latency compared to RocksDB with traditional fair sharing. In summary, our key contributions are:


\begin{itemize}
    \item We introduce \newRFfull, a new family of algorithms for sharing resources with high preemption delays, that explicitly bound client wait times to a configurable parameter \newRFsymbol while ensuring high utilization.
    
    \item We design algorithms that provide both \newRFsymbol-fairness and \newRFsymbol-Pareto-efficiency for two key resources that primarily cause tail-latency spikes in storage systems, write buffers and read caches, based on the dynamics of each resource.
    
    \item We implement our algorithms in \sys, an extension of RocksDB, and show that \sys bounds end-to-end tail-latency spikes of well-behaved clients (to \newRFsymbol time units) in the presence of high-demand workloads.
\end{itemize}

%% file: 02-background-osdi26.tex
\section{Motivation}
\label{sec:bg-and-mot}


In this section, we summarize why performance isolation is critical for cloud storage systems before highlighting the limitations of fair sharing algorithms: they do not account for the high preemption delays of resources like buffers and caches.



\subsection{Isolation from High-Demand Workloads}

In the cloud, storage systems offer strict latency SLOs to well-behaved clients that remain within their provisioned demand. To meet such SLOs cost-effectively, storage systems multiplex physical cloud resources and support diverse clients across multiple tenants. However, extremely high-demand workloads are unavoidable in the cloud; such workloads can arise from application heterogeneity, large legitimate workloads, unintended misconfigurations and bugs, or targeted malicious behavior~\cite{cheng2018analyzing, reiss2012heterogeneity, yang2021large, cloudflare2022_isolation, borthakur2022_bursty_traffic}. These workloads can drive request rates far beyond what storage systems are designed to support. Isolating well-behaved clients from throughput degradation or tail-latency spikes that cause SLO violations is a fundamental challenge in cloud storage systems. Our experiments show that well-behaved clients running standard YCSB load-A workloads on RocksDB, insert-heavy workloads on PostgreSQL, or sequential read-only fio workloads on the Linux OS page cache, experience large tail-latency spikes in the presence of high-demand workloads, as shown in \autoref{tab:intro-mot}. Most cloud storage systems adopt rate limits~\cite{databricksRateLimiting}, but rate limits alone cannot provide strong performance isolation for well-behaved clients. 

\subsection{Fair Sharing of Resources}
Fair sharing has been a standard approach to achieve performance isolation and high resource utilization in other domains.
It assigns each client \changebars{its \textit{fair}}{a fair} share of a resource and redistributes unused capacity to clients with high demand. Specifically, it introduces two fairness properties:

\noindent \textbf{Share guarantee}: Each of $n$ clients receives at least $1/n^{\text{th}}$ of the shared resource, regardless of other clients' demands. 

\noindent \textbf{Pareto efficiency}: The resource share of one client cannot be increased without decreasing the share of another client.

Fair sharing is widely adopted for CPU scheduling~\cite{drf, drfq}, cluster scheduling~\cite{drf, vuppalapati2023karma}, and networking~\cite{processorsharing, fair_without_trade, demers1989analysis}, where it provides performance isolation via the share guarantee property and achieves high utilization through Pareto efficiency. In these domains, fair sharing enables clients to quickly converge to their fair resource shares and allocates unused resources to clients with high demand. Fundamentally, these resources can be rapidly redistributed among clients, i.e., they have negligible preemption delays. For example, link bandwidth can be reassigned within a few microseconds once the current packet is transmitted. Subsequent work extending fair sharing (e.g., across time, multiple resources, or with shared data and workloads, as outlined in \S\ref{sec:related-work}) similarly assumes resources with negligible preemption delays.

\subsection{Resources with High Preemption Delays}

Storage systems break traditional fair-sharing assumptions: storage resources cannot be quickly reassigned among clients because they exhibit high preemption delays. Write buffers and read caches in databases and file systems force long delays on clients that attempt to acquire fair shares of these resources, since reclaiming them requires expensive disk I/O operations. Freeing up buffer space involves flushing multiple dirty pages to disk, where each flush is roughly 100$\times$ slower than an in-memory write. Similarly, reclaiming cache space requires fetching entries from disk that were previously cached but have been evicted by other clients. These preemption delays can be arbitrarily long, depending on system events and background tasks such as compaction and garbage collection. These also compete for the underlying physical resources, like the available disk bandwidth. In practice, storage systems such as RocksDB and PostgreSQL—and the OS page cache underlying them—have resources with high preemption delays.

Fair sharing is thus only realized \textit{eventually} in storage systems, because resource allocations converge to fair shares only after the necessary flushes and fetches complete. As shown in \autoref{fig:example}(a), the well-behaved client, attempting to reclaim their fair share at $T0$, experiences a long delay ($T1-T0$ time units) before receiving its fair share due to the resource's high preemption delay, potentially violating its latency SLOs.


This is not just a theoretical concern. We see this pattern in RocksDB, one of the most used key-value stores today. We evaluate the interference from a single high-demand writer on well-behaved writers. Each client runs standard YCSB Load-A workloads. Well-behaved clients consistently operate below their provisioned demand, 
and the high-demand client's peak demand can saturate the system. We compare three resource allocation policies: (1) the default FCFS (first-come-first-serve) policy in RocksDB, (2) static per-client resource quotas, and (3) fair sharing. Compared to static quotas, fair sharing increases the overall write throughput by redistributing spare capacity; however, it does so at the cost of tail latency. As shown in \autoref{tab:intro-mot}, well-behaved clients experience more than 200$\times$ worse tail latency with fair sharing 
in the presence of a single high demand. Therefore, performance isolation remains a key challenge despite fair sharing resources in storage systems.


%% file: 03-fairness-osdi26.tex
\section{\propfull}
\label{sec:rad_definition}

In this section, we introduce, and define the desired properties of, \newRFfull, a new family of algorithms that are designed to share resources with high preemption delays.

\newRFfull satisfies two key properties: \newRFsymbol-fairness, defined to guarantee performance isolation for resources with high preemption delays, and \newRFsymbol-Pareto-efficiency, which ensures high utilization of such resources. These properties must hold regardless of other clients’ demands, so we define them under \textit{\worstcasedemand} conditions:

\vspace{5pt}
\noindent
\textbf{\newRFsymbol-fairness:}
When a client demands resource shares at time $t$, it must receive these shares, up to its fair share by time $t + \newRFsymbol$, i.e., within a maximum delay (\newRFsymbol) even under the \worstcasedemand.

\vspace{5pt}
\noindent
\textbf{\newRFsymbol-Pareto-efficiency:}
The resource share of one client cannot be increased without either decreasing the share of another client or causing another client to violate the maximum delay (\newRFsymbol) under the \worstcasedemand.
\vspace{5pt}

\newRFsymbol-fairness introduces the maximum delay (\newRFsymbol)\footnote{Here, we discuss a single global maximum delay (\newRFsymbol) for all clients; however, this parameter can trivially extend to per-client delays, allowing systems to distinguish well-behaved, latency-sensitive clients (with a smaller \newRFsymbol).}, and bounds client delays to acquire their fair share of resources.
It states that clients can experience a delay of no more than \newRFsymbol time units to receive their fair share of a resource regardless of other clients’ demands, as long as their demand conforms to the assumed \worstcasedemand conditions. Under these \worstcasedemand conditions, \newRFsymbol-Pareto-efficiency states that any unused resource units not necessary to satisfy \newRFsymbol-fairness should be redistributed to clients with unmet demand.

When \newRFsymbol is set to zero, clients must receive their fair shares immediately, and \newRFfull reduces to an isolated system that assigns each client its fair share of resources (static quotas). If \newRFsymbol is unbounded ($\newRFsymbol=\infty$), \newRFfull includes traditional fair sharing. In practice, system administrators or designers configure \newRFsymbol, similar to SLOs and rate limits in commercial storage systems~\cite{databricksRateLimiting}.
\changebars{}{\devbali{only if k = n right? I feel like we should introduce k here at a high level, like system administrators can make assumptions about bursts}}

The \newRFsymbol-fairness property composes cleanly across multiple resources within a system, even when each resource has a high preemption delay, and bounds the end-to-end tail latency spike of well-behaved clients. If a client requires several resources, it incurs no more than \newRFsymbol\ time units to obtain its fair share of each of these resources. If $R$ resources are acquired sequentially, the client must receive its fair share of each resource within $R\!\times\!\newRFsymbol$ time units, and for resources acquired in parallel, its fair share of each resource can be reclaimed independently within \newRFsymbol\ time units.

More generally, when multiple systems, each enforcing its own \newRFsymbol$_s$ (i.e., satisfying \newRFsymbol-fairness under the same \worstcasedemand conditions), are stacked together, the end-to-end tail latency spikes of well-behaved clients are bounded by:
\[
\newRFsymbol_{total} = 
\begin{cases}
\sum_{s} \newRFsymbol_s, & \text{sequential access} \\
\max_{s} \newRFsymbol_s, & \text{parallel access}
\end{cases}
\]

\paragraph{Resource reclamation.} Importantly,  \newRFsymbol-fairness constrains only the maximum delay (\newRFsymbol) before a client regains its fair share of resources, it does not prescribe how reclamation is carried out within that delay (\newRFsymbol), simplifying the design of \newRFfull algorithms (\S\ref{sec:minimal-reservations}). Reclamation timelines may, however, impact other metrics such as average latency or throughput. Several reclamation timelines are possible: a client may receive no share until $t+\newRFsymbol$ and then its full share at once (all-at-once), receive a gradually increasing share over time (linear), or receive an immediate partial share with the remainder granted by $t+\newRFsymbol$ (stepwise). 
We only define \newRFsymbol-fairness with respect to the absolute time it takes to get to the fair share, to give administrators a simple and easy-to-understand property about how clients can affect each other.

\section{Designing \newRFfull Algorithms}
\label{sec:minimal-reservations}

In this section, we discuss the key mechanism underlying the design of our \newRFfull algorithms (\S\ref{sec:policies-rocksdb}) and the \worstcasedemand conditions that our algorithms assume.

When a client requests its fair share of a resource, the algorithm can allocate unused resources or reclaim resources that are currently in use by other clients. If in-use resources must be reclaimed and the resource's preemption delay is less than the configured maximum delay ($<$\newRFsymbol), then \newRFsymbol-fairness is trivially achieved. However, if reclaiming the resource takes longer than the maximum delay ($>$\newRFsymbol), then the algorithm must withhold some amount of the resource from distribution to other clients, even when they are unused, which we term resource \textit{reservations}, to provide \newRFsymbol-fairness.

\newRFfull algorithms compute the required reservation sizes to bound client delays to receive their fair share to \newRFsymbol time units (providing \newRFsymbol-fairness) and redistribute unused resources (achieving \newRFsymbol-Pareto-efficiency and high utilization).

\begin{figure}[t]
    \includegraphics[width=0.99\linewidth]{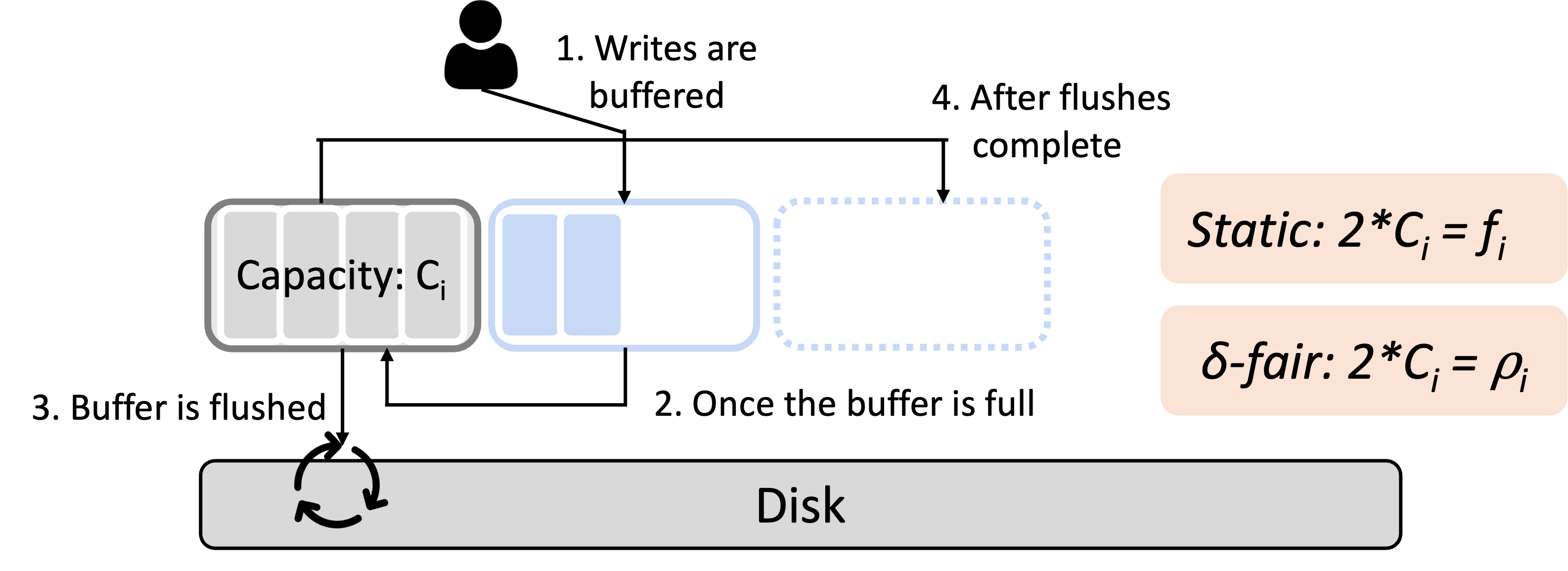}
    \caption{Delayed fair sharing for write buffer. \(f_i\) is a client's fair share, \(C_i\) is a portion of buffer capacity, and \(\rho_i\) is the reservation size.}
    \label{fig:example1}
\end{figure}


\paragraph{Algorithm.}
Consider a client that writes to a toy write buffer with a fair share of \(f_i = 100~\mathrm{MB}\). Once the buffer reaches a certain capacity \(C_i\)—here, 50\% of the client’s fair share (\(C_i = 50~\mathrm{MB}\)) as shown in \autoref{fig:example1}—the buffer starts flushing dirty pages to disk in blocks of size \(B = 64~\mathrm{kB}\) and replenishes its capacity.

With an algorithm that assigns static quotas to clients, well-behaved clients that stay below their provisioned demand do not fill up their fair share \(f_i\). However, high-demand clients fill their \(f_i\), causing their writes to queue up. Such high-demand clients cannot use the use the spare capacity from well-behaved clients. Since traditional fair-sharing algorithms are work-conserving and eagerly assign unused capacity from well-behaved clients to high-demand clients, high-demand clients use up all the available unused capacity causing well-behaved clients to experience large tail latency spikes when they ramp up to their fair share (\autoref{tab:intro-mot}, \S\ref{sec:bg-and-mot}).

\newRFfull algorithms, by contrast, compute a reservation \(\rho_i\) for well-behaved clients and redistribute only the remaining capacity to other high-demand clients. They guarantee that well-behaved clients can reclaim this remaining capacity within \newRFsymbol time units, even under \worstcasedemand. For instance, when \newRFsymbol is configured to 200~ms and the minimum flush rate of the client under  \worstcasedemand is \(R_\mathrm{min} = 50~\mathrm{MB/s}\), a client can reclaim up to 
\(R_\mathrm{min} \times \newRFsymbol = 10~\mathrm{MB}\) during that period. The remaining 90~MB is reserved for the client:
\(\rho_i = f_i - (R_\mathrm{min} \times \newRFsymbol) = 100 - 10 = 90~\mathrm{MB}\), even if it is unused.

Concretely, \newRFfull algorithms first estimate the minimum refill rate \(R_\mathrm{min}\) of the resource under the assumed \worstcasedemand conditions. They then compute the maximum amount of the resource that can be safely reclaimed in \newRFsymbol time units (\(R_\mathrm{min} \times \newRFsymbol\)) and allow only this amount to be redistributed. The remainder is reserved for the client to preserve \newRFsymbol-fairness, as outlined in \autoref{algorithm}, satisfying both \newRFsymbol-fairness and \newRFsymbol-Pareto-efficiency.


\paragraph{The \worstcasedemand conditions.}
Estimating $R_\mathrm{min}$, and thus computing the required reservation \(\rho_i\) requires understanding the \worstcasedemand. While multiple models are possible, in this paper, we use a simple model based on the maximum number of clients that can concurrently ramp up to their fair shares.


\changebars{}{
As shown in \autoref{fig:example}(b), the \newRFfull algorithm achieves this via resource reservations. 
\natacha{There's some repetition here: I'd go straight to "To bound client delays while achieving high utilization, our \newRFfull algorithm uses resource reservations} A small portion of the resource (e.g., the write buffer) is reserved (in grey) such that the well-behaved client is promised immediate access to it. As a result, the high-demand client can flush fewer dirty pages, and the well-behaved client experiences no more than \newRFsymbol time units of delay in receiving the remaining portion of its fair share. The reserved portion is withheld from redistribution to the high-demand client even when the well-behaved client is not using it (before time $T0$ and after $T2$). However, the unreserved portion is redistributed to the high-demand client to achieve high utilization and \newRFsymbol-Pareto-efficiency. Our \newRFfull algorithms compute the required reservation sizes that meet both the properties.
\natacha{I'm confused, you talk to me about an algorithm, but then you start with an example. Could you switch? Also, you don't actually explain the algorith - paragraph very confusing to me}
\natacha{I'm very confused as to the role of 4.1 and the link with 4.2. You talk about 4.1 as introducing an algorithm, but you don't actually describe the algorithm, instead it's mostly through an ad-hoc example. Then you talk about how to configure \peakdemandthreshold, but you don't link it back to the algorithm. Same comment as in 3, it would help flow a lot if you didn't have two subsections, and instead had: 1) an explanation of the algorithm 2) how you do minimal reservations with respect to a workload and how you compute k, and then only then give the example. What you're discussing is not complex enough that it warrants so much space. Makes the section feel empty}\soujanya{working on the missing example}
}

\changebars{}{
\soujanya{TODO for @Tyler: Pseudo algorithm goes here! Here's a brief outline of the explanation:

1. Our algorithms model the preemption delay of a resource ($\tau$)
2. Then we model the minimum refill rate of the resource under the \worstcasedemand
3. Then we estimate the maximum amount of resource that can be reclaimed within \newRFsymbol time units at minimum refill rate (max. reclaimed; we need a term for this)
4. And we determine the reservation size to be (fair share - maximum reclaimable)
5. If ($\tau > \newRFsymbol$), then we make a reservation of size (fair share - maximum reclaimed  > 0), else we do not need any reservation.
6. Our algorithms redistribute unused resource shares (available resource units - reservation size) to other clients with high demand

Qn: where do we talk about accounting for background system events: GC or compactions? refill rate or maxreclaimed?

\soujanya{Add the relevant information about reservation sizes and tradeoffs; see the comment below.}}
}


\begin{algorithm}[t]
\caption{ComputeReservations:}
\begin{enumerate}
    \item Requires: Clients, fair shares \(F\), and minimum refill rate under worst-case demand \(R_{min}\)
    \item For each client $c \in$ Clients:
    \begin{enumerate}
        \item maxReclaimed[$c$] $=$ \(R_{\min} \times \newRFsymbol\)
        \item $\rho$[$c$] $= \max(F[c] - \text{maxReclaimed}[c], 0)$
    \end{enumerate}
    \item Return $\rho$
\end{enumerate}
\label{algorithm}
\end{algorithm}

The \textit{\peakdemandthreshold} ($k$), denotes the maximum number of clients that may simultaneously reclaim their fair share. It helps us estimate the \emph{minimum refill rate}, i.e., the rate at which a resource can be released when $k$ clients request their fair share concurrently.
It represents the \worstcasedemand: even if fewer than $k$ clients demand their fair share currently, our algorithms make larger reservations, which lower utilization, in favor of preserving \newRFsymbol-fairness. Furthermore, if all clients can demand their fair share at once ($k=n$), we may need static quotas to achieve \newRFsymbol-fairness.
Our algorithms derive the \peakdemandthreshold from historical traces. Fortunately, real-world traces show that $k$ is often small; typically, in traces with 64 clients~\cite{snowflake-nsdi20, yang2021large, cao2020characterizing}, only about 8–10\% of the active clients have simultaneous demand bursts. More complex models can be explored to capture \worstcasedemand and \newRFfull naturally extends to such \worstcasedemand conditions.

\section{\newRFfull in \sys}
\label{sec:policies-rocksdb}


We now derive \newRFfull allocation policies that compute and implement reservations in RocksDB for two of its resources: the write buffer and the read cache. Our policies compute each resource's preemption delay, minimum refill rate, and the minimal reservation size required to satisfy the two properties: \newRFsymbol-fairness and \newRFsymbol-Pareto-efficiency, under the \worstcasedemand ($k$). \changebars{While we choose RocksDB as a representative system, \newRFfull broadly applies to storage systems and resources with high preemption delays.}{ RocksDB is a representative system that enables discussing resource- and system-dependent factors to computing reservation sizes, however, \newRFfull applies to other storage systems and resources with high preemption delays.}

\begin{figure}[t]
    \centering
    \includegraphics[width=0.85\linewidth]{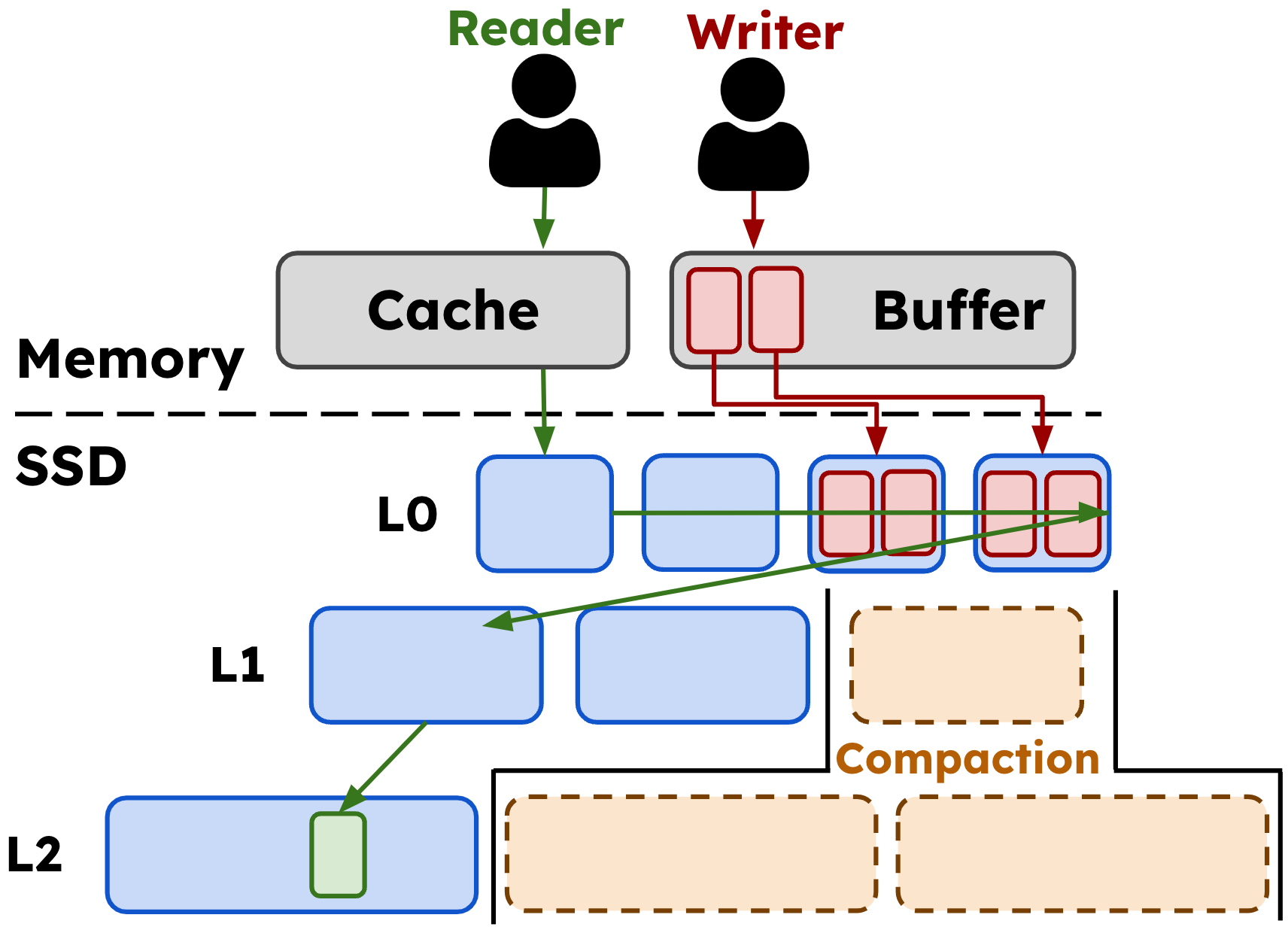}
    \caption{\textbf{RocksDB Architecture Overview.}}
    \label{fig:rocksdb-overview}
\end{figure}

We build \sys to extend RocksDB with \newRFfull.
\sys has multiple storage resources: the write buffer (or memtable), write-ahead log, the LSM tree, read cache, and bloom filters, as shown in \autoref{fig:rocksdb-overview}. We target a standard deployment of multi-tenant RocksDB that supports clients with independent workloads (no shared datasets or applications). Consequently, clients need not share the write-ahead logs, LSM trees, or the bloom filters (via per-client column families). The remaining shared resources: write buffer and read cache, are the primary contributors to tail latency spikes. Other physical resources, such as CPU and read/write I/O bandwidth (with negligible preemption delays), can be quickly reassigned, so conventional fair sharing suffices. 

\begin{table}[t]
  \centering
  \begin{tabularx}{\columnwidth}{@{}>{\hspace{0pt}}l X@{}}
    \toprule
    \textbf{Notation} & \textbf{Meaning} \\ \midrule
    \rowcolor{gray!15}
    \newRFsymbol        & \newRFsymbol–fairness bound \\
    \rowcolor{gray!15}
    $k$                 & \peakdemandthreshold \\
    \midrule
    \rowcolor{blue!15}
    $n$                 & Number of clients \\
    \rowcolor{blue!15}
    $f_i$               & Client $i$’s fair share \\
    \rowcolor{blue!15}
    $U_i$               & Client $i$’s usage \\
    \rowcolor{blue!15}
    $R_{\mathrm{read}}$ & Read I/O bandwidth capacity \\
    \rowcolor{blue!15}
    $R_{\mathrm{write}}$& Write I/O bandwidth capacity \\
    \rowcolor{blue!15}
    $B$                 & Write buffer segment size \\
    \rowcolor{red!15}
    Res.\ pool          & Reservation pool \\
    \rowcolor{red!15}
    Global pool         & Global pool \\
    \bottomrule
  \end{tabularx}

  \caption{Key parameters in our policies.  
    The fundamental maximum delay (\newRFsymbol) parameter, the \peakdemandthreshold ($k$) which captures \worstcasedemand, and the remaining derived system parameters (in blue) and computed parameters (in red).}
  \label{tab:notation-config}
\end{table}

\begin{figure*}[t]
    \centering
    \begin{minipage}[b]{0.48\linewidth}
        \flushright
        \includegraphics[width=0.85\linewidth]{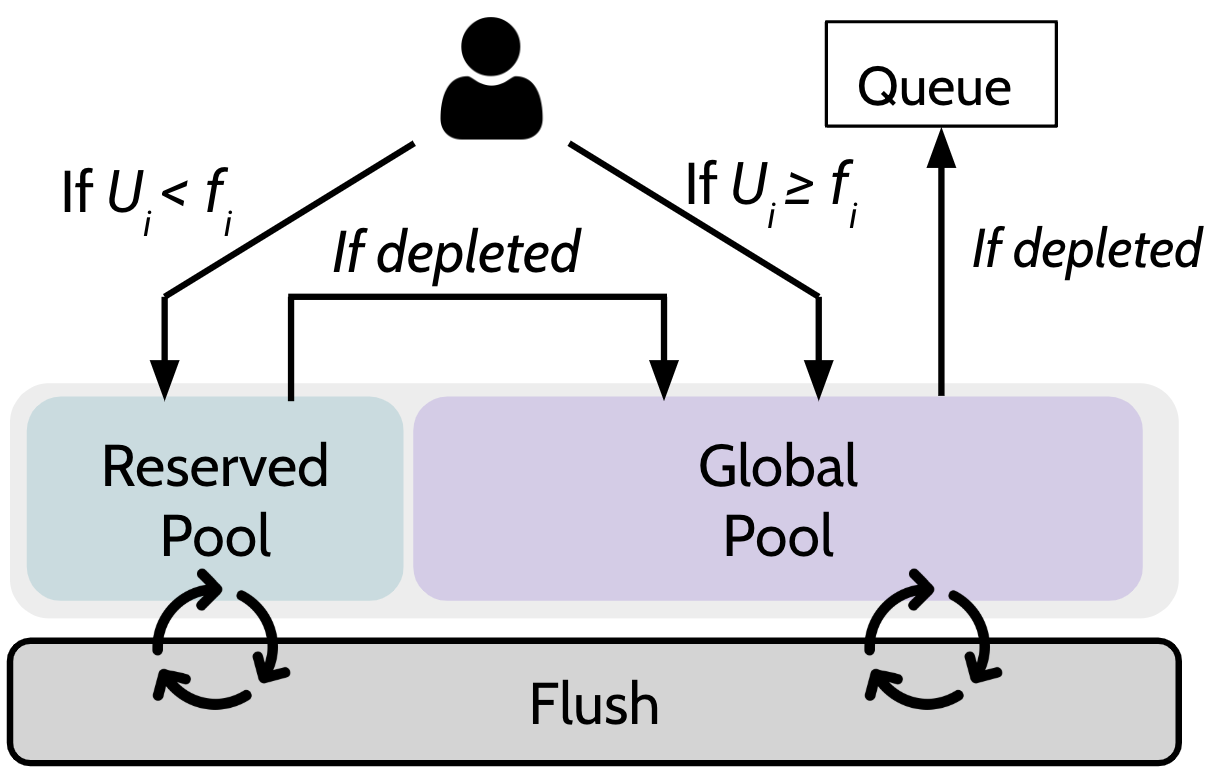}
        \caption{\textbf{Write Buffer Manager (WBM).} A reserved pool ensures bounded reclamation delay; excess usage draws from a global pool.}
        \label{fig:wbm-diagram}
    \end{minipage}
    \hfill
    \begin{minipage}[b]{0.48\linewidth}
        \flushleft
        \includegraphics[width=0.85\linewidth]{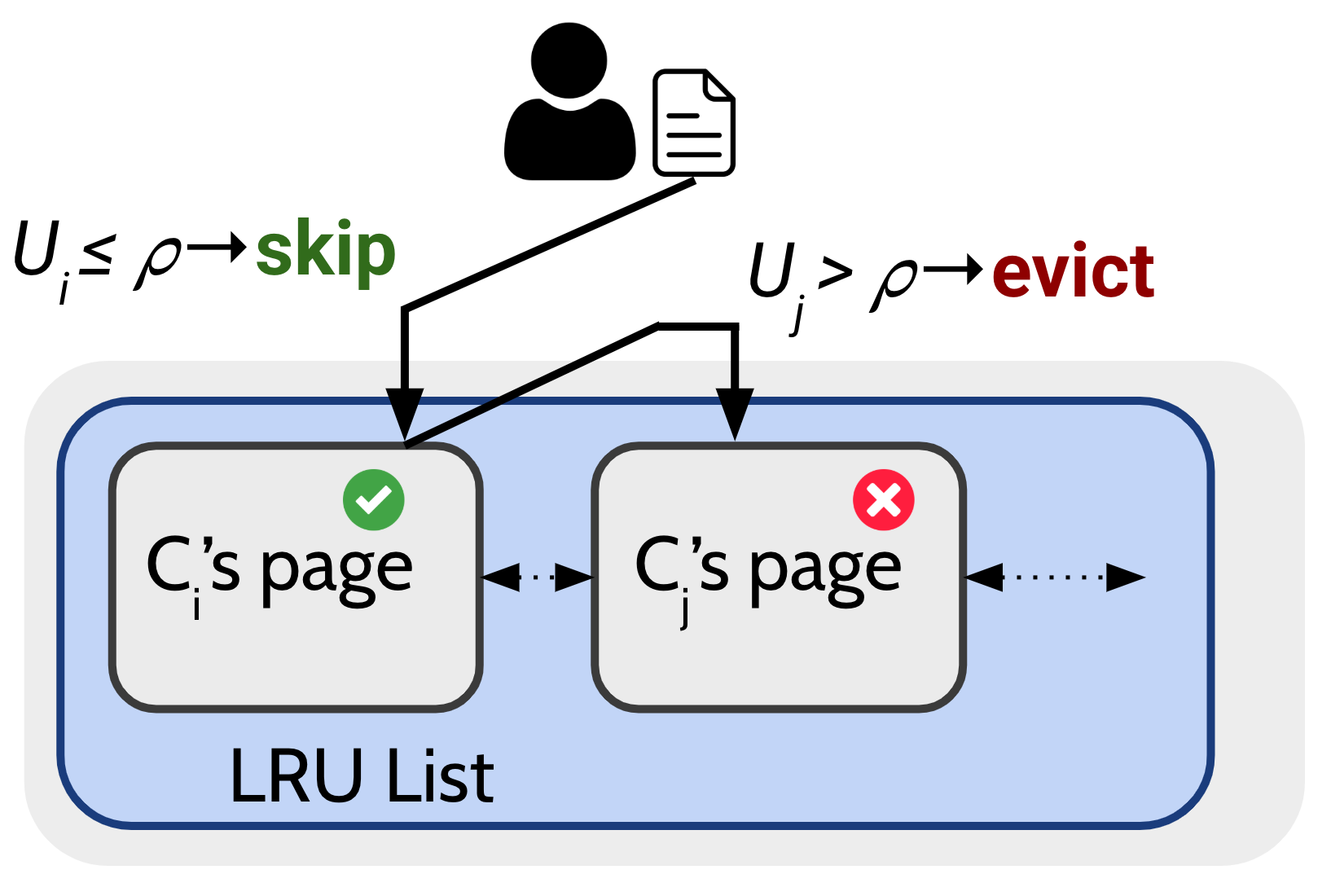}
        \caption{\textbf{Cache Manager (CM).} Per‑client reservations prevent eviction below \(\resCache\); capacity above \(\resCache\) remains shareable.}
        \label{fig:cache-diagram}
    \end{minipage}
\end{figure*}

We now present the \newRFfull policies—first for the write buffer, then for the read cache. In these policies, \newRFsymbol is the fundamental tunable maximum delay, and $k$ specifies the \worstcasedemand model. All remaining quantities (as summarized in \autoref{tab:notation-config}) are either system-derived parameters or values computed by the policy.

\subsection{\newRFsymbol-Fair Write Buffer Policy}
\label{sec:buffer-policy}

We present a \newRFsymbol-fair policy that computes reservations for the write buffer and discuss an allocation policy that enforces these reservations and achieves \newRFsymbol-Pareto-efficiency.

\subsubsection{Minimal buffer reservations}
\label{subsec:buffer-minres}

\changebars{
We analyze a single writer (i) with a fair share \(f_i\) and compute its reservation size $\resBuffer^{(i)}$; we omit the sub- and super-script in \S\ref{subsec:buffer-minres} for readability. The delay stems from clients having to flush segments to disk before relinquishing them.
}{
We analyze a single client \(i\) with fair buffer share \(f_i\) (measured in bytes) and a reserved buffer space of size \(\resBuffer^{(i)}\). For readability, we write \(f = f_i\) and \(\resBuffer = \resBuffer^{(i)}\) for this client and drop the sub- and super-script. The delay in the write buffer stems from the client having to flush buffered segments to disk before relinquishing them.
}

Let \(B\) denote the flush granularity (i.e., the write-buffer segment size in bytes) and \(R_{\mathrm{write}}\) the total disk write bandwidth available for flushing write-buffer segments. Then, \(\lceil f/B\rceil\) is the number of segments in the client’s fair share. 
Flushing \(\lceil f/B\rceil\) segments takes:
\[
T_{\mathrm{flush}}
= \Big\lceil \frac{f}{B} \Big\rceil \frac{B}{R_{\mathrm{write}}}.
\]

If \(T_{\mathrm{flush}} \le \newRFsymbol\), no reservation is needed because the client's full fair share can be flushed and reallocated within the maximum delay $\newRFsymbol$. Otherwise, \(\resBuffer\) bytes are reserved such that the \emph{residual} demand \((f-\resBuffer)\) can be reclaimed within \newRFsymbol:
\[
\Big\lceil \frac{f-\resBuffer}{B} \Big\rceil \frac{B}{R_{\mathrm{write}}} \le \newRFsymbol.
\]

Under \worstcasedemand, up to \(k\) clients may attempt to reclaim their fair share simultaneously. In that case, each client effectively receives at most \(R_{\mathrm{write}}/k\) bandwidth, so the minimal refill rate per client is \(R_{\mathrm{write}}/k\). Reclaiming the residual \((f-\resBuffer)\) for a single client then takes:
\[
T_{\mathrm{flush\mbox{-}multi}}
= \Big\lceil \frac{f-\resBuffer}{B} \Big\rceil \frac{B}{R_{\mathrm{write}}/k},
\]
and the minimal reservation size for this client solves:
\begin{equation}
  \min \ \resBuffer
  \quad\text{s.t.}\quad
  \Big\lceil \frac{f-\resBuffer}{B} \Big\rceil \frac{B}{R_{\mathrm{write}}/k} \le \newRFsymbol.
  \label{eq:buffer-reservation-calc}
\end{equation}
Equivalently, each client can reclaim at most \((R_{\mathrm{write}}\newRFsymbol)/k\) bytes within \newRFsymbol; we choose \(\resBuffer\) so that \(f-\resBuffer\) does not exceed this bound. This derivation applies directly under weighted fairness by interpreting \(f_i\) as the weight-proportional fair share of the buffer for client \(i\). 

\changebars{
For \(n\) clients, we set the total reserved pool size as the sum of per-client reservations we obtain from \eqref{eq:buffer-reservation-calc}, \(\resBuffer = \sum_{i=1}^n \resBuffer^{(i)}\); the next subsection describes how the buffer allocation policy implements this as a single shared reserved pool.

}{
\soujanya{
So far, we've discussed the reservation for a single client; we could discuss what the reservation would look like across clients (total: $n$)?} \tyler{Took a simple first stab below, but could expand on it.}

For \(n\) clients, we obtain a per-client reservation \(\resBuffer^{(i)}\) from \eqref{eq:buffer-reservation-calc} and set the total reserved pool size to \(\sum_{i=1}^n \resBuffer^{(i)}\); the next subsection describes how we implement this as a single shared reserved pool.
}

\subsubsection{Buffer allocation policy}
\label{subsec:buffer-policy}

\autoref{fig:wbm-diagram} illustrates the allocation policy used for the write buffer. The policy manages two logical pools of the total write buffer capacity \(C\): a \emph{reserved pool} and a \emph{global pool}. The reserved pool aggregates per-client reservations and has the total capacity \(\resBuffer\). The global pool has capacity \(G = C - \resBuffer\) and holds the remaining buffer space that can be shared elastically among clients. The reserved pool is a single shared pool and is not partitioned per client. Only clients currently below their fair share may draw from it.

Let \(U_i\) be the current buffer usage of client \(i\) across both the reserved and global pools. On a write by client \(i\), the policy proceeds as follows:
\begin{enumerate}[leftmargin=*]
  \item If \(U_i < f_i\) (client \(i\) is below its fair share), allocate the segment from the reserved pool if it has free space; if the reserved pool is depleted, allocate from the global pool; if both pools are depleted, queue the request.
  \item If \(U_i \ge f_i\) (client \(i\) is at or above its fair share), allocate only from the global pool; if the global pool has no free space, queue the request.
\end{enumerate}
When a segment is freed after a flush, it first refills the reserved pool (until its capacity \(\resBuffer\)); any remaining freed capacity is returned to the global pool. Queued requests are drained in increasing order of a client's utilization \(U_i/f_i\), so that clients furthest from their fair share receive priority when space becomes available.  Reservation sizes follow \eqref{eq:buffer-reservation-calc} and thus adapt automatically to system rates and the configured \newRFsymbol\ (and the optional \(k\)).


\subsection{\newRFsymbol-Fair Read Cache Policy}
\label{sec:cache-policy}

\changebars{We next derive the reservation size for the read cache, where eviction is fast but refilling the cache with data from disk is slow.}{
We next apply the same \newRFsymbol-fair model \natacha{what's a model here? First time you use the word I believe}   to the read cache, where eviction is fast but refilling from disk is slow.} Unlike write-buffer segments, cache pages are associated with particular tenants and are therefore not interchangeable across clients without a refill delay (evicting the page and reading a new one from disk). To bound per-client delays to \newRFsymbol time units, we compute minimal per-client cache reservations, which are the minimal cache size thresholds below which a client's cache pages cannot be evicted. The remaining cache capacity above these thresholds is elastic and can be shared across all clients. Similar to the write buffer policy, the reservation is computed based on the configured \newRFsymbol (and $k$) and the time it takes to read evicted cache pages from disk.

\subsubsection{Minimal cache reservations}
\label{subsec:cache-reservations}

Eviction itself is instantaneous, so delay stems from reading pages from disk. We analyze a single client \(i\) with a fair cache share \(f_i\) (in bytes) and a reserved cache size threshold of \(\resCache^{(i)}\). For readability, we write \(f = f_i\) and \(\resCache = \resCache^{(i)}\) for this client and drop the sub- and superscript in \S\ref{subsec:cache-reservations}.

Let \(R_{\mathrm{read}}\) denote the total read bandwidth available for serving cache misses, and let  \(\mathrm{AMP}\) be the read-amplification factor: refilling one byte of cached data requires reading \(\mathrm{AMP}\) bytes from disk. The effective refill bandwidth for this client is therefore \(R_{\mathrm{read}}/\mathrm{AMP}\). If the client temporarily loses all bytes cached above \(\resCache\) by eviction, the residual demand that must be refilled upon the client requesting its fair share is \((f - \resCache)\) bytes. Refilling this residual within delay \(\newRFsymbol\) requires:
\[
(f - \resCache)\,\frac{\mathrm{AMP}}{R_{\mathrm{read}}} \le \newRFsymbol
\quad\Rightarrow\quad
\resCache \ge f - \newRFsymbol\!\left(\frac{R_{\mathrm{read}}}{\mathrm{AMP}}\right).
\]

Under \worstcasedemand, up to \(k\) clients may attempt to reclaim their fair share of the cache simultaneously. In that case, each client effectively receives at most \(R_{\mathrm{read}}/k\) read bandwidth, giving an effective refill bandwidth of \(R_{\mathrm{read}}/(k \cdot \mathrm{AMP})\). Refilling the residual \((f - \resCache)\) for a single client then takes:
\[
T_{\mathrm{refill\mbox{-}multi}}
= (f - \resCache)\,\frac{\mathrm{AMP}}{R_{\mathrm{read}}/k},
\]
and the minimal reservation size for this client solves:
\begin{equation}
  \min \ \resCache
  \quad\text{s.t.}\quad
  (f - \resCache)\,\frac{\mathrm{AMP}}{R_{\mathrm{read}}/k} \le \newRFsymbol.
  \label{eq:cache-res-min-r}
\end{equation}
Equivalently, each client can refill at most \(\big(R_{\mathrm{read}}\newRFsymbol\big)/(k \cdot \mathrm{AMP})\) bytes within \newRFsymbol; we choose \(\resCache\) so that \(f-\resCache\) does not exceed this bound.  As in the write buffer, weighted fairness is obtained by interpreting \(f_i\) as the weight-proportional fair share of cache for client \(i\) and solving \eqref{eq:cache-res-min-r} per client.

\subsubsection{Cache allocation and eviction}
\label{subsec:cache-alloc}

\autoref{fig:cache-diagram} illustrates the allocation policy used for the read cache. We overlay reservation enforcement on the cache’s existing eviction policy (LRU in RocksDB). Unlike the write buffer, there is no global pool of available cache pages: every cached page holds data for a particular tenant, and pages or reservations are not fungible across clients. The cache maintains a per-client size thresholds \(\resCache^{(i)}\) obtained from \eqref{eq:cache-res-min-r}. When admitting a new page on a cache miss, the eviction walk skips candidate victim pages whose owner is at or below its threshold \(\resCache^{(i)}\) and evicts the first page owned by a client whose usage exceeds its limit. Thus, each client $i$ retains at least \(\resCache^{(i)}\) bytes of cache, bounding its refill delay by \newRFsymbol and providing \newRFsymbol-fairness, while any cache capacity above these thresholds remains elastic and is dynamically redistributed to other clients with unmet demand, providing \newRFsymbol-Pareto-efficiency (\S\ref{ssec:eval-micro:cache}).

%% file: 06-impl.tex
\section{\sys Implementation}


We implement \sys as an extension of RocksDB. 
Our modifications to RocksDB fall within two categories: (a) implementing policies that adhere to the properties of \newRFfull (\S\ref{sec:policies-rocksdb}), and (b) enabling per-client resource accounting. They are primarily in the write buffer management, block cache, I/O rate limiter, and stall trigger components. 


\changebars{To enforce per-client usage, \sys}{
We modify RocksDB's existing managers to implement our per-resource \newRFsymbol-adhering scheduling policies rather than implementing a single, centralized resource scheduler. 

To implement \newRFsymbol-fair policies and enforce per-client usage, \sys traces resource usage back to clients. It} 
assigns each client an ID and propagates the client ID through RocksDB’s request handling path, to ensure that background tasks (compaction and flushes) are also accounted to the correct client.
RocksDB's resource managers for write buffers, cache, and I/O use this ID to maintain per-client usage statistics.
\changebars{}{
\sys extends the built-in I/O rate limiter that manages the read and write I/O rates separately. }
We build on this to implement a token-based fair I/O scheduler inspired by Gimbal~\cite{min2021gimbal}. 
\changebars{}{
Each client is guaranteed a base allocation of write bandwidth (its fair share) and unused capacity is redistributed to clients with higher demand, ensuring high utilization.
}
\changebars{
We modify the Write Buffer Manager (WBM), to intercept requests prior to buffering them and track the following statistics: (a) per-client write buffer usage, (b) global pool availability, (c) reserved pool availability, and (d) currently queued write requests, and handles a write request (as detailed in \S\ref{subsec:buffer-policy}).
}{
We modify the Write Buffer Manager (WBM), which intercepts requests prior to buffering them, to track the aggregate memory usage across all column families. We can then implement and enforce our \newRFsymbol-fair write buffer policy by prevent eviction for clients under their reserved share. 
The WBM object tracks statistics: (a) per-client write buffer usage, (b) global pool availability, (c) reserved pool availability, and (d) currently queued write requests, and then handle a write request (as detailed in \S\ref{sec:buffer-policy}).
}


\sys builds on the sharded LRU Cache Manager (CM) to manage the read cache. It still uses a single logical cache as in RocksDB, but extends RocksDB's CM to track cache usage per client and maintain per-client reservation sizes. A read request that is a cache hit only updates LRU statistics and does not pass through CM. Instead, the CM intercepts reads when they bring pages in from the disk on a cache miss. When a page added to the cache, the CM identifies the shard it belongs to and navigates the LRU list to find the first block belonging to a client that is above its reservation size threshold, as detailed in  (\S\ref{subsec:cache-reservations}). 


\changebars{\sys sets aside approximately 30\% of the total I/O bandwidth available for compactions, similar to production-grade LSM-based KV stores~\cite{cockroachcloud}. This prevents compactions from disrupting client performance.
}{
Compactions consume significant I/O resources for reading, merging, and then writing large files to disk. They interfere with the I/O bandwidth otherwise used to reclaim read cache or write buffer resources promptly. 
To prevent compactions from disrupting client performance, we follow the approach adopted by production LSM-based KV stores~\cite{cockroachcloud} and set aside approximately 30\% of the total I/O bandwidth available for compactions. This dedicated allocation ensures that compactions proceed at a steady, controlled pace without monopolizing I/O resources. Within the dedicated compaction bandwidth, the same fair \io scheduling policies are applied.



}
Similarly, to prevent a single high-demand client from compromising performance isolation, \sys modifies global stall triggers in RocksDB to apply on a per-tenant basis. It tags stalls with the client who generates them. When a client's stall threshold is exceeded, only that client’s requests are throttled, leaving others unaffected.

%% file: 07-eval.tex
\section{Evaluation}
\label{sec:evaluation}

We evaluate the end-to-end benefits of \newRFfull and show how \sys isolates well-behaved clients from extremely high-demand workloads (\S\ref{ssec:eval-e2e-ycsb}). We trace these benefits back to each \newRFsymbol-fair policy (\S\ref{ssec:eval-micro-all}) and discuss their composability across resources (\S\ref{ssec:eval-micro:multires}). Finally, we highlight the adaptability of our policies to different maximum delay (\newRFsymbol) and \peakdemandthreshold ($k$) configurations (\S\ref{ssec:eval-sensitivity}).


\noindent \textbf{Experimental Setup}.
Our macrobenchmarks are run on a Debian \texttt{n2-standard-96} VM (96 vCPUs, 48 physical cores, 384 GB RAM) on GCP, with 16×375 GB NVMe local SSDs (6,240 MB/s read, 3,120 MB/s write). The storage system and the workload generation are pinned to separate NUMA nodes. 
Our microbenchmarks are run on a smaller GCP-VM with 980 MB/s write, and 1280 MB/s read, \io bandwidth.

\noindent \textbf{Baselines}.
We compare \sys against two baselines. The primary baseline, RocksDB with instantaneous fair sharing implemented on each resource, represents the latest research on fair resource allocation, and RocksDB with static per-client resource quotas reflects common industry practices. These baselines also capture the two ends of the isolation–utilization trade-off spectrum.

\noindent \textbf{RocksDB Configuration}.
We follow the official RocksDB tuning guide~\cite{rocksdb_tuning_guide}. We disable the OS page cache and use per-client column families, where clients do not share data or workloads—a standard setup in cloud deployments. Their workload configurations and \sys-specific parameters (\newRFsymbol and $k$) are detailed within each experiment. Across all RocksDB baselines and \sys, clients fairly share \io bandwidth. Because I/O bandwidth has a negligible preemption delay ($<$1 ms), we adopt a simple token-based fair I/O sharing algorithm. To avoid interference from compactions, we dedicate about 30\% of the total I/O bandwidth for compactions, similar to other key-value stores in production~\cite{cockroachcloud}. This dedicated allocation ensures that compactions proceed at a steady, controlled pace without monopolizing I/O resources; within this reserved bandwidth, the same fair \io sharing policies are applied. Finally, to further reduce performance interference, we modify global write stalls to affect only the column family that triggers them across all baselines.

\noindent \textbf{Clients}.
Clients  are pinned to the remote NUMA node and run in an open-loop setting, issuing requests into per-client queues which are serviced by RocksDB/\sys worker threads. Request rates, access patterns, and working set sizes are detailed within each experiment. Due to the lack of publicly available multi-tenant storage traces (to the best of our knowledge), we construct representative workloads from analyzing caching traces from Twitter~\cite{yang2021large} and Snowflake~\cite{snowflake-nsdi20}; we describe them next and plan to open-source these workloads.

\noindent \textbf{Multi-tenant workload generation}. 
In each experiment, we use the Yahoo Cloud Serving Benchmark (YCSB) suite~\cite{ycsbpaper} to generate client workloads and derive per-client request inter-arrival times and read/write distributions from production traces~\cite{snowflake-nsdi20, yang2021large, cao2020characterizing}. Our trace analysis shows that only 8–10\% of clients ramp up demand concurrently; we use this to configure \sys's \peakdemandthreshold ($k$) in each experiment. 

\noindent \textbf{Measurements}.
We report an average of 3 runs where each run takes between two and ten minutes, depending on the experiment. For evaluations using a cache, we warm-up the cache and only begin reporting statistics after throughput convergence.

\subsection{YCSB Macrobenchmark}
\label{ssec:eval-e2e-ycsb}

We use a multi-tenant YCSB macrobenchmark to evaluate \sys's end-to-end performance under concurrent read and write contention. \sys's \newRFsymbol-fair policies reduce well-behaved clients' P99 latency by up to 9.3$\times$ compared to instantaneous fair sharing, while preserving high utilization. Overall throughput is within 4\% of fair sharing and up to 30\% higher than static quotas. 

In the macrobenchmark, there are $n=$64 clients, and the workload comprises  28\% read-only, 13\% write-only, and 59\% mixed read-write clients, resulting in 20 read-only, 10 write-only, and 34 mixed clients; read-only clients are evenly split between YCSB-C and YCSB-E, write-only clients use Load-A and YCSB-A (5 each), and mixed clients use YCSB-B, D, and F (11, 11, and 12 clients, respectively), with request rates uniformly scaled to induce maximum load on a shared RocksDB instance. Each client operates on a 40 GB dataset (1M records, 4 kB in size) with the default Zipfian access pattern. There are two high-demand clients that cause simultaneous contention on both the write- and read-serving resources: (a) a writer, which issues a write that can fill up the entire write buffer every 30 seconds, and (b) a reader, which issues high-rate random reads for a 30s duration, every 2 minutes. 
For \sys, we set $\newRFsymbol_{buffer}$=350ms, and $\newRFsymbol_{cache}$=250ms, and based on the trace analysis we set $k=$6 ($\approx9$\% of 64).

\begin{figure}[t]
    \centering
    \includegraphics[width=\columnwidth]{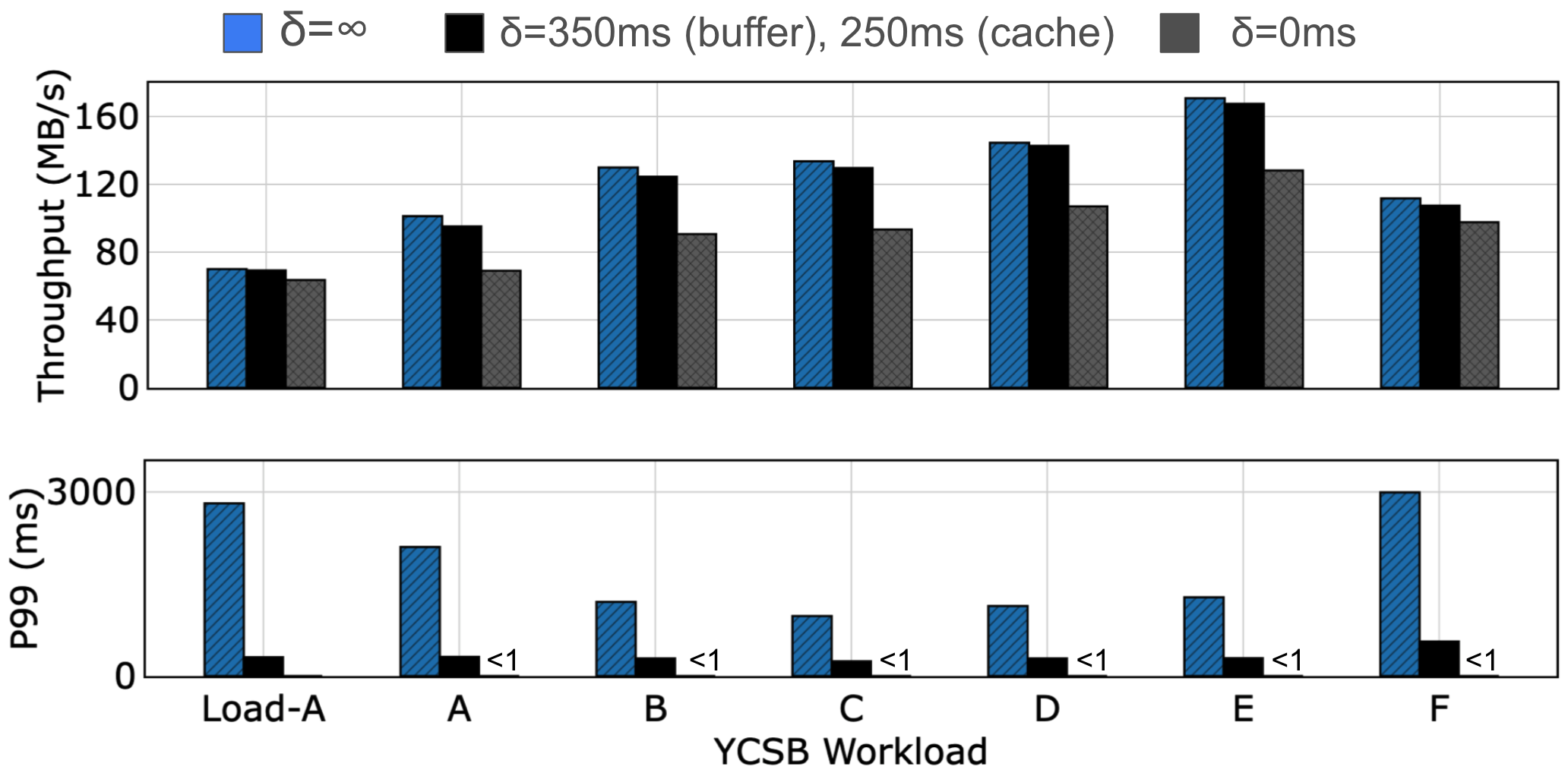}
    \caption{\sys on YCSB Macrobenchmark. Index (from left to right): \newRFsymbol = \{inf, $\newRFsymbol_{buffer}=$350ms, $\newRFsymbol_{cache}=$250ms, 0\}. We show the throughput and p99 latency of clients per YCSB workload. The \newRFsymbol =inf (blue) depicts RocksDB with fair sharing and \newRFsymbol =0 represents static quotas (gray). \newRFsymbol-fair policies (black) achieve up to 9$\times$ lower p99 latency compared to fair sharing with comparable overall throughput.}
    \label{fig:ycsb_macro}
\end{figure}


We report the throughput and P99 latency per YCSB workload, as shown in Figure~\ref{fig:ycsb_macro}, and summarize the overall system throughput in Table~\ref{tbl:ycsb_macro}. \sys delivers strong performance isolation: it consistently improves tail latency from \textbf{4$\times$} by up to \textbf{9.3$\times$} compared to RocksDB with instantaneous fair sharing. \sys also maintains high resource utilization, achieving up to \textbf{30\%} higher throughput than static reservations and only \textbf{4\%} below that of instantaneous fair sharing, due to resource reservations.
From our observations, write-heavy workloads have higher P99 latency than read-heavy workloads. Clients running workload F (read-modify-write) exhibit the highest P99 latency from sequentially accessing the read cache and the write buffer: the delays in receiving cache shares, reading and modifying the data, and then receiving write buffer shares, add up as explained later in \S\ref{ssec:eval-micro:multires}. Notably, all P99 latencies remain within the configured maximum delay (\newRFsymbol), showcasing the accuracy of our \newRFsymbol-fair policy design~\S\ref{sec:policies-rocksdb}. 

For clients running YCSB-E (scan-intensive), despite reading more data, the P99 latency was not significantly higher than that of clients with other read-heavy workloads. We observe this because both scan-intensive and other read-heavy clients operate under a large backlog of pending requests in both scenarios. Finally, when well-behaved clients issue requests well below their fair shares, \sys can further increase utilization relative to RocksDB with static quotas: \newRFsymbol-fair reservations provision only enough capacity to satisfy the configured maximum delay (\newRFsymbol), allowing unused reserved capacity to be reclaimed by other clients and boosting throughput by up to 30\%.

\begin{table}[]
\small
\begin{tabular}{@{}cccc@{}}
\toprule
\textbf{FairDB}        & \textbf{Read (GB/s)} & \textbf{Write (GB/s)} & \textbf{Overall (GB/s)} \\ \midrule
$\delta$=0 & 5.4     & 7.5       & 6.2 (\textbf{1$\times$})   \\ 
$\delta$ = inf  & 7.4     & 0.91  & 8.3 (\textbf{1.34$\times$}) \\
\begin{tabular}[c]{@{}c@{}} $\delta_{buffer}=350ms,$\\ $\delta_{cache}=250ms$\end{tabular} & 7.1 & 0.88 & 8.0 (\textbf{1.29$\times$}) \\
\bottomrule
\end{tabular}
\caption{\sys overall combined throughput: \sys with \newRFsymbol-fair policies has comparable throughput relative to RocksDB with static quotas (\newRFsymbol = 0), or fair sharing (\newRFsymbol = inf). \sys (\newRFsymbol = 600ms) achieves \textbf{4--9.3$\times$} better tail latency compared to fair sharing.}
\label{tbl:ycsb_macro}
\end{table}

\subsection{Microbenchmarks}
\label{ssec:eval-micro-all}

We break down the end-to-end improvements of \sys and trace them back to each of our \newRFsymbol-fair policies for the write buffer (\S\ref{ssec:eval-micro:buffer}) and the read cache (\S\ref{ssec:eval-micro:cache}). We show that our \newRFsymbol-fair policies isolate well-behaved clients from bursty, high-demand ones, while redistributing spare capacity for greater system throughput.
The microbenchmarks are run on a smaller GCP-VM (to reduce cost) with 980 MB/s write,and 1280 MB/s read \io bandwidth. We discuss the performance of clients running write-only YCSB Load-A and read-only Run-C workloads.


\changebars{
}{
\subsubsection{\io Bandwidth.}
\label{ssec:eval-micro:io}

\sys fairly shares \io bandwidth to prevent significant P99 latency spikes and throughput degradation from \io bandwidth contention.
A reading client's cache misses compete for read \io bandwidth, allowing a high-demand client to dominate the \io bandwidth share with frequent fetches. Conversely, writing clients' buffer flushes demand write \io bandwidth, allowing a high-demand writer to dominate the \io share with frequent flushes. 
Due to the negligible preemption delay ($<$1ms) of \io bandwidth, \sys simply implements a token-based fair I/O sharing algorithm. In all our experiments, fair \io sharing is enabled among clients across all RocksDB baselines.
\soujanya{No numbers in this subsection at all? We should have it as a paragraph if we are not discussing any results}
}

\begin{figure}[ht]
    \centering
    \includegraphics[width=\columnwidth]{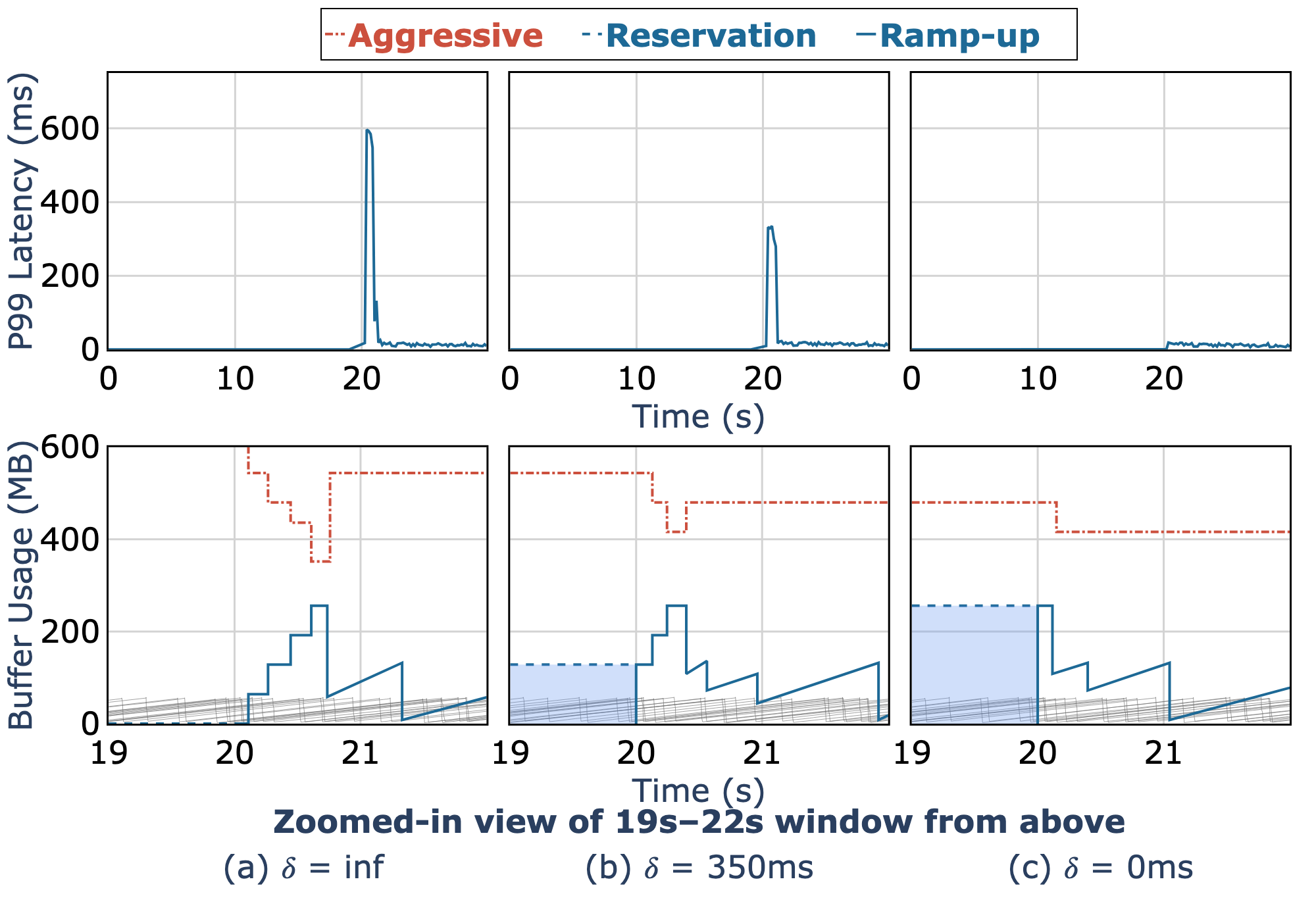}
    \caption{Write buffer microbenchmarks. We show the p99 latency and buffer usage of the ramp up client (blue), aggressive (red), and well-behaved clients (grey). \sys enforces minimal reservations to effectively meet ramp-up clients' maximum delay (here, \newRFsymbol = 350ms)}
    \label{fig:buffer_microbenchmark}
\end{figure}

\subsubsection{Write Buffer.}
\label{ssec:eval-micro:buffer}

The write-buffer microbenchmark evaluates how \sys's \newRFsymbol-fair policy bounds well-behaved writers' tail latency while preserving buffer utilization. Relative to RocksDB with fair sharing, \sys provides a \textit{configurable} reduction in tail latency spikes, as seen in Figure~\ref{fig:buffer_microbenchmark}. \sys effectively reserves buffer capacity based on clients' configured maximum delays and supports $\newRFsymbol_{buffer}=0$ms for strict latency sensitivity, $\newRFsymbol_{buffer}=\inf$ for high maximum delay, and intermediate values such as 350ms; other \newRFsymbol settings are explored in \S\ref{ssec:eval-sensitivity}. 

We use 16 clients: 12 well-behaved writers at 50~MB/s each, 2 high-demand writers at 192~MB/s, and 2 well-behaved clients that ramp up later in the experiment. The 14 always-active clients collectively issue around 980~MB/s of writes and saturate the available write bandwidth (980~MB/s). The total write buffer capacity is 2~GB, resulting in a per-client fair share of 2~GB / 16 = 128~MB. At 20~seconds, the 2 ramp-up clients begin ramping up: each first issues a batch write of size 128~MB, equal to its fair share of the buffer, and then continues writing steadily at 50~MB/s, matching the other well-behaved clients. All clients run the YCSB Load-A workload (write-only) with 4~kB requests. In this microbenchmark, we set $k=2$, corresponding to the maximum number of well-behaved clients that may ramp up simultaneously.

Under instantaneous fair sharing ($\newRFsymbol_{buffer}=\inf$), the ramp-up clients' batch writes cannot complete until the high-demand clients release 2$\times$128~MB = 256~MB of buffer space. Given that the high-demand clients jointly receive 380~MB/s of the write bandwidth (980~MB/s minus 12$\times$50~MB/s from well-behaved writers), they are expected to take 256~MB / 380~MB/s $\approx$ 673~ms to flush enough data. In our experiment, this takes 596~ms (11\% below the theoretical estimate). When $\newRFsymbol_{buffer}=0$, \sys behaves like static quotas: it fully reserves the ramp-up clients' fair shares, and they do not experience any tail latency spikes. For an intermediate configuration of $\newRFsymbol_{buffer}=350$ms, \sys bounds the ramp-up clients' P99 delay to 332~ms, as shown in Figure~\ref{fig:buffer_microbenchmark}, effectively enforcing their configured maximum delay.

With the ramp-up clients' maximum delay configured at $\newRFsymbol_{buffer}=350$ms and $k=2$, \sys reserves 128~MB of buffer capacity (6\% of the 2~GB buffer) for potential ramp-up demand. A bound of 350~ms is 350/673 $\approx$ 52\% of the worst-case delay under instantaneous fair sharing, which suggests that roughly 48\% of the backlog must be hidden by reservations. Because flushing occurs at the granularity of 64~MB buffer pages, the reservation size is rounded up to two 64~MB pages (128~MB), per the ceiling operator in~\autoref{eq:buffer-reservation-calc}. As a result, during ramp-up, only the remaining 128~MB must be flushed, which takes about 332~ms (5\% below the configured $\newRFsymbol_{buffer}$). For the stricter setting $\newRFsymbol_{buffer}=0$, \sys reserves the full additional demand of the $k=2$ ramp-up clients (2$\times$128~MB = 256~MB, 12.5\% of the total buffer capacity), allowing them to instantly ramp up to their fair shares without preemption delay. Thus, \sys's write-buffer reservations provide configurable performance isolation while minimally impacting buffer utilization.

\subsubsection{Read Cache.}
\label{ssec:eval-micro:cache}

\begin{figure}[t] \centering \includegraphics[width=\columnwidth]{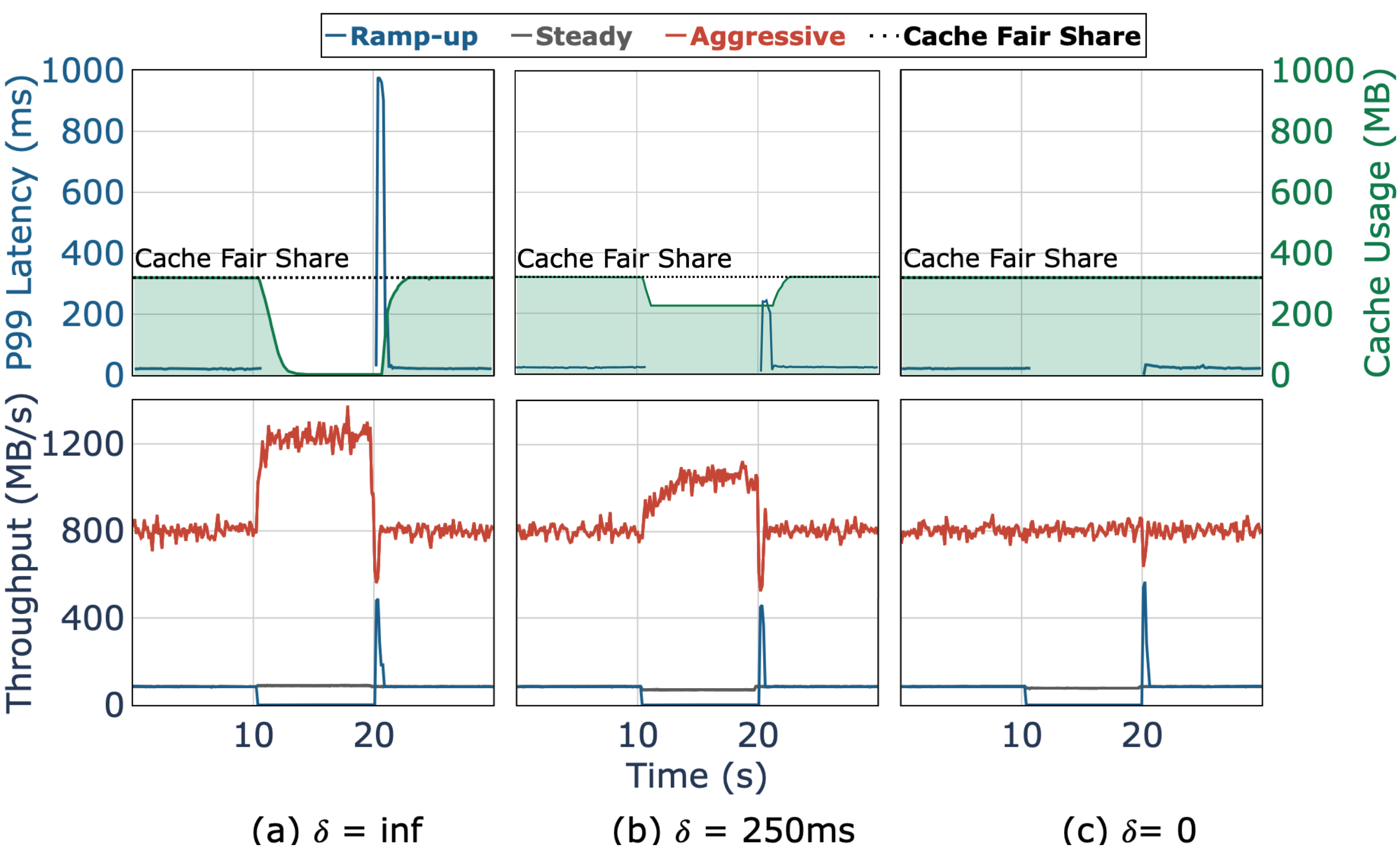} \caption{Cache microbenchmarks. We show the p99 latency and throughput of ramp-up (blue), aggressive (red), and well-behaved (grey) clients. \sys has minimal reservations to effectively meet ramp-up clients' maximum delay (here, \newRFsymbol = 500ms)} \label{fig:cache_microbenchmark} \end{figure}

We next evaluate \sys's \newRFsymbol-fair read-cache policy using a microbenchmark analogous to the write-buffer experiment in \S\ref{ssec:eval-micro:buffer}. The goal is to measure ramping-up well-behaved readers' tail latency and ensuring that unused cache space is reclaimed to achieve higher throughput. Relative to RocksDB with fair sharing, \sys provides a configurable reduction in tail latency and improves throughput when clients tolerate higher maximum delays. In this experiment, we focus on a moderate setting of $\newRFsymbol_{cache}=500$ms; other \newRFsymbol settings are explored in \S\ref{ssec:eval-sensitivity}.

We use 32 readers: 30 well-behaved clients that read steadily at 50~MB/s each, and 2 high-demand clients that read at 1400~MB/s with a 75\% cache miss rate. We use more readers here than writers in \S\ref{ssec:eval-micro:buffer}, reflecting our trace analysis of production workloads~\cite{snowflake-nsdi20,yang2021large}. The total cache size is 10~GB, resulting in a per-client fair share of 10~GB / 32 = 320~MB. Each well-behaved client has a 320~MB working set (equal to its fair share), so all of its requests can be served from cache, while each high-demand client has a 4$\times$ larger working set. All 32 clients are active for most of the run and collectively issue 4300~MB/s of reads, generating 2100~MB/s of disk bandwidth demand, which saturates the available 1280~MB/s. Clients run YCSB Run-C (read-only) with 4~kB records and the default Zipfian distribution.

A single well-behaved client goes offline at 10~seconds and begins ramping up again at 20~seconds. To reclaim its cache share, it issues a 320~MB read (implemented as a \texttt{multi-get}) and then continues reading steadily at 320~MB/s. Under instantaneous fair sharing ($\newRFsymbol_{cache}=\inf$), this ramp-up client shares the refill bandwidth with the 2 high-demand clients. Given 1280~MB/s of total disk bandwidth and 3 clients contending for cache refills, the ramp-up client receives about 320~MB/s, so refilling its 320~MB working set takes roughly 1~second, as seen in Figure~\ref{fig:cache_microbenchmark}. When $\newRFsymbol_{cache}=0$, \sys behaves like static quotas: it fully preserves the client's cache share while it is idle, so the ramp-up client can immediately resume with no latency spike. For an intermediate configuration of $\newRFsymbol_{cache}=250$ms, \sys bounds the client's P99 ramp-up delay to 241~ms, effectively enforcing its configured maximum delay.

With the ramp-up client's maximum delay set to $\newRFsymbol_{cache}=250$ms, \sys reserves 240~MB of cache for this client while it is offline, so its utilization drops only to this reservation size. As a result, the ramp-up client needs to refill only the remaining 80~MB of its 320~MB working set, which takes about 241~ms instead of 1~second. From 10--20~seconds, the reserved space (240~MB) remains with the offline client, but the remaining cache capacity is fully used by the high-demand clients. This improves cache utilization: high-demand client throughput increases by 48\%, and overall system throughput increases by 29\% compared to static quotas. Thus, \sys's read-cache reservations provide configurable performance isolation while allowing unused cache capacity to be reclaimed for higher throughput.

\subsection{Delays Compose across Multiple Resources}
\label{ssec:eval-micro:multires}

We next evaluate how \sys's \newRFsymbol-fair policies compose when clients contend for both cache and buffer resources, demonstrating that per-resource delay bounds translate into an end-to-end bound for multi-resource operations. 

\begin{figure}[t]
    \centering
    \begin{subfigure}[t]{\columnwidth}
        \centering
        \includegraphics[width=\linewidth]{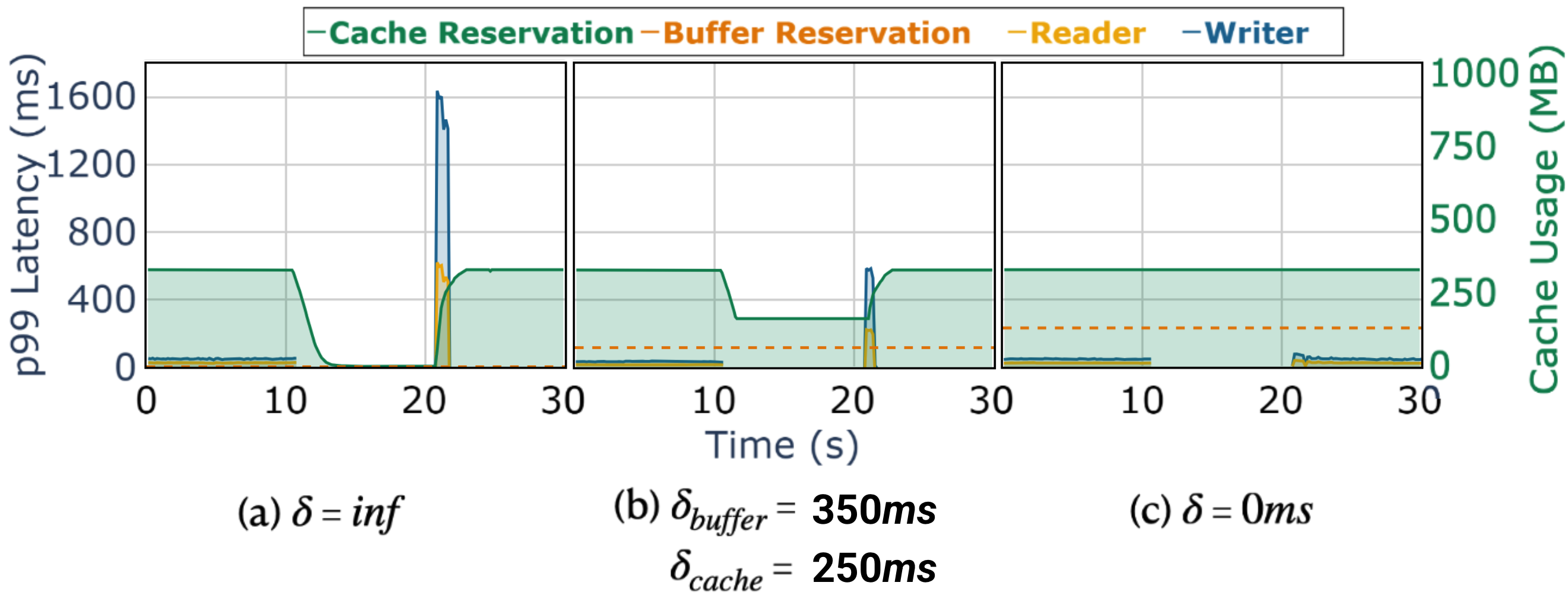}
        \caption{The additive composition of delays in \sys when resources are acquired sequentially. Clients have read-modify-write workloads, YCSB Run-F.}
        \label{fig:multi_microbenchmark_additive}
    \end{subfigure}
    

    \begin{subfigure}[t]{\columnwidth}
        \centering
        \includegraphics[width=\linewidth]{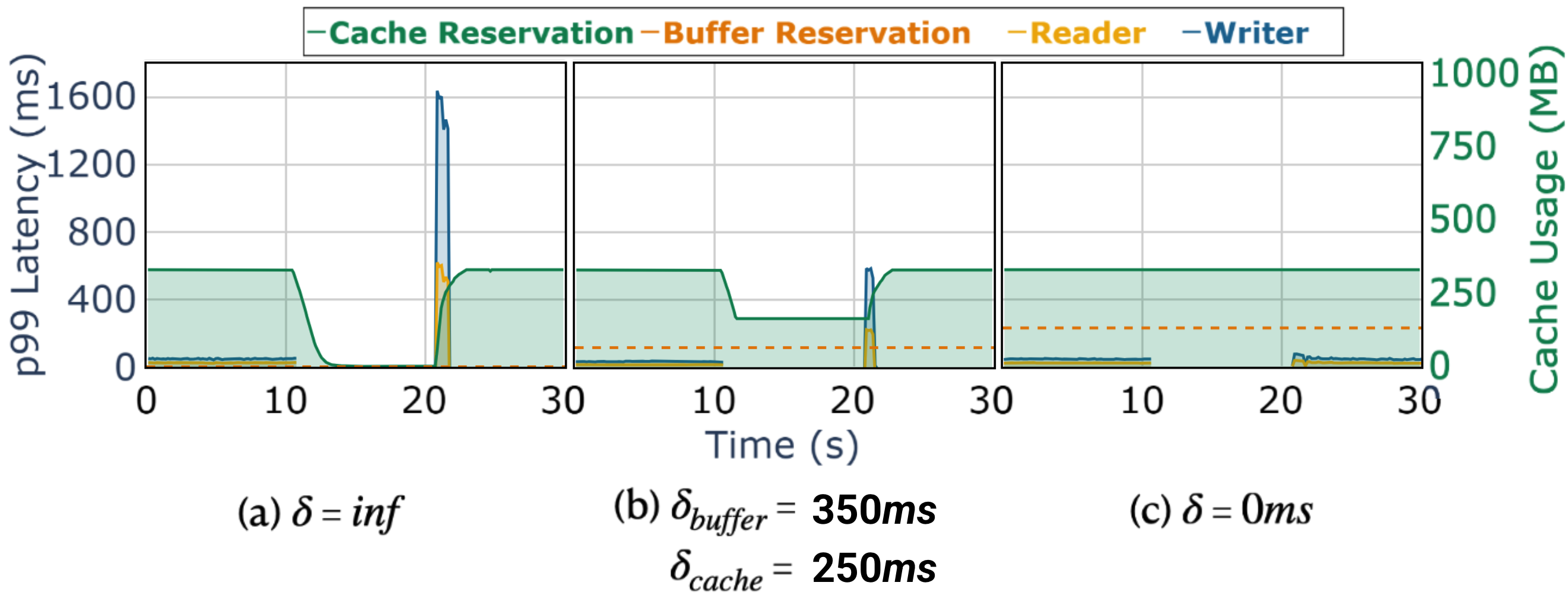}
        \caption{The best-case latency spike in \sys \ie the maximum delay across resources when they are acquired in parallel. Clients have independent read and write requests, YCSB Run-A.}
        \label{fig:multi_microbenchmark_max}
    \end{subfigure}

    \caption{Multiple resources. The delay bounds compose across multiple resources for the end-to-end latency spikes of ramp-up clients; \sys meets the configured maximum delays for each resource.}
    \label{fig:multi_microbenchmark}
\end{figure}

To evaluate the composability of delays in \sys, we use clients that demand both read caches and write buffers: they perform \texttt{read-modify-write} and mixed reads and writes, with YCSB Run-F and Run-A workloads, respectively. We fair share \io across clients and enable our \newRFsymbol-fair policies for caches and buffers. For explainability, we combine our experiment settings from the two per-resource microbenchmarks (\S\ref{ssec:eval-micro-all}) into a single multi-resource microbenchmark. There are 32 total clients, with 16 readers and 16 that have a mixed read-write workload (either Run-F or Run-A). There are 2 aggressive clients, each of which is an aggressive reader (1400 MB/s) and aggressive writer (192 MB/s). The well-behaved clients read at 50 MB/s and write at 50 MB/s. The cache capacity is 10GB, with a per-client fair share of 320MB. The buffer capacity is 4GB, which a per-client fair share of 128MB.

We evaluate three settings, and \sys allocates per-client reservations based on the configured latency sensitivity of each setting: ($\text{\newRFsymbol}_{\text{buffer}},\text{\newRFsymbol}_{\text{cache}}=$0ms) for strict latency sensitivity, ($\text{\newRFsymbol}_{\text{buffer}},\text{\newRFsymbol}_{\text{cache}}=\inf$) for high maximum delay, and ($\text{\newRFsymbol}_{\text{buffer}}=350$ms, $\text{\newRFsymbol}_{\text{cache}}=250$ms) for moderate maximum delay.


~\autoref{fig:multi_microbenchmark_additive} shows a p99 latency of the ramp-up client of 565ms: 221ms from the read and 344ms from the subsequent write. In \sys, the client's read-modify-write operation, individually meet the configured delay bounds for buffer ($\text{\newRFsymbol}_{\text{buffer}}=350$ms) and cache ($\text{\newRFsymbol}_{\text{cache}}=250$ms). The ramp-up client reads at 50 MB/s for the first 10s, goes offline, and begins ramping up at 20 seconds. During ramp-up, it first reads 320 MB, then modifies and writes 128 MB of the read records. This amounts to the client demanding its fair share of cache and buffer sequentially, the worst case for tail latency.  In \sys, the end-to-end additional resource acquisition delay is bounded by the sum of configured per-resource \newRFsymbol bounds. In RocksDB with fair sharing (i.e., \newRFsymbol=$inf$), clients experience a 1639ms spike, and with static quotes (i.e., \newRFsymbol=0), clients experience no delays.

When resources are acquired in parallel, the delay bounds compose as a \textit{maximum} of the per-resource delays. To demonstrate this, the client submits both a large read (multi-get) and batch write request to the system at the same time, and we report the P99 latency of completing both requests in ~\autoref{fig:multi_microbenchmark_max}. With$\text{\newRFsymbol}_{\text{buffer}}=350$ms and $\text{\newRFsymbol}_{\text{cache}}=250$ms, the p99 latency of the client is 338ms due to the preemption delay of the write buffer. The delay in acquiring cache resources (236ms) is incurred in parallel and therefore does not contribute to the p99 latency. This demonstrates the best-case additive \newRFsymbol delay composition in \sys.

\subsection{Sensitivity and Configurability (\newRFsymbol, $k$)}
\label{ssec:eval-sensitivity}

We next study how \sys behaves as operators tune its configuration knobs: the per-resource delay bounds (\newRFsymbol) and the \peakdemandthreshold $k$. Across all settings, our \newRFsymbol-fair policies keep clients' P99 latencies within the configured delay bounds under contention, while exposing a clear trade-off: relaxing \newRFsymbol or choosing a smaller $k$ reduces reservation sizes and increases throughput, whereas tighter bounds or larger $k$ provide stronger isolation at the cost of additional reserved capacity.

\subsubsection{Latency Sensitivity}

We first vary the per-resource delay bounds and measure the impact on reservations, P99 latency, and throughput, reported in \autoref{tbl:delta-reservations}. In all cases, the observed additional P99 delay remains at or below the configured \newRFsymbol bound. For the read cache, increasing \newRFsymbol from 0~ms to $inf$ reduces the fraction of cache reserved for ramp-up clients and increases system throughput from 3.10~GB/s to 3.92~GB/s, while the additional P99 delay grows from $<\!1$~ms to 977~ms. For the write buffer, increasing \(\newRFsymbol_{buffer}\) similarly shrinks reservation sizes from 12.5\% to 0\% and increases the tolerated additional delay from $<\!1$~ms to 596~ms. These trends illustrate the per-resource trade-off between isolation and utilization: looser \newRFsymbol bounds require fewer reservations and yield higher throughput, while tighter bounds reserve more capacity to bound tail latency.

We next examine how these bounds compose across multiple resources using the multi-resource microbenchmark from \S\ref{ssec:eval-micro:multires}. \autoref{tab:multires-delta-sweep} sweeps combinations of buffer and cache bounds, with $(\newRFsymbol_{\mathrm{buffer}}, \newRFsymbol_{\mathrm{cache}}) \in \{200,350,500\}\times\{250,500,750\}$, and reports the end-to-end P99 latency of a read–modify–write operation. In every configuration, the observed P99 latency is below the sum \(\newRFsymbol_{\mathrm{total}} = \newRFsymbol_{\mathrm{buffer}} + \newRFsymbol_{\mathrm{cache}}\). For example, with \(\newRFsymbol_{\mathrm{buffer}}=500\)~ms and \(\newRFsymbol_{\mathrm{cache}}=250\)~ms, the configured bound is 750~ms and the measured P99 latency is 741~ms. This confirms that \sys’s per-resource \newRFsymbol-fair policies compose into a predictable end-to-end bound for multi-resource operations, while still allowing operators to independently tune the bounds for the cache and buffer.


\begin{table*}[t]
  \centering
  \begin{tabular}{@{}cccccc@{}}
    \toprule
    \textbf{Read Cache (\newRFsymbol)} & \textbf{0 ms} & \textbf{250 ms} & \textbf{500 ms} & \textbf{750 ms} & \textbf{inf} \\
    \midrule
    Additional Delay (ms) & <1 ms & 236 & 484 & 729 & 977 \\
    System Throughput (GB/s)    & 3.10 (+0 \%) & 3.36 (+8.4 \%) & 3.64 (+17.4 \%) & 3.82 (+23.2 \%) & 3.92 (+26.5 \%) \\
    \midrule
    \textbf{Write Buffer (\newRFsymbol)} & \textbf{0 ms} & \textbf{200 ms} & \textbf{350 ms} & \textbf{500 ms} & \textbf{inf} \\
    \midrule
    Additional Delay (ms) & <1 ms & 161 & 332 & 494 & 596 \\
    Reservation Size (\%) & 12.5 \% & 9.4 \% & 6.3 \% & 3.1 \% & 0 \% \\
    \bottomrule
  \end{tabular}
  \caption{Sensitivity to \newRFsymbol bounds for single resources. \sys keeps the additional P99 delay within the configured \newRFsymbol while reducing reservation sizes and increasing throughput as the bound is relaxed for both the read cache and the write buffer.}
  \label{tbl:delta-reservations}
\end{table*}

\begin{table}[t]
  \centering
  \small
  \begin{tabular}{@{}cccc@{}}
    \toprule
    \textbf{$\newRFsymbol_{\mathrm{buffer}}$ (ms)} & 
    \textbf{$\newRFsymbol_{\mathrm{cache}}$ (ms)} &
    \textbf{$\newRFsymbol_{\mathrm{total}}$ (ms)} &
    \textbf{P99 Latency (ms)} \\
    \midrule
    200 & 250 & 450 & 425 $\pm$ 6 \\
    200 & 500 & 700 & 684 $\pm$ 4 \\
    200 & 750 & 950 & 925 $\pm$ 8 \\
    \midrule
    350 & 250 & 600 & 582 $\pm$ 6 \\
    350 & 500 & 850 & 831 $\pm$ 3 \\
    350 & 750 & 1100 & 1040 $\pm$ 9\\
    \midrule
    500 & 250 & 750 & 741 $\pm$ 3 \\
    500 & 500 & 1000 & 983 $\pm$ 7 \\
    500 & 750 & 1250 & 1154 $\pm$ 9 \\
    \bottomrule
  \end{tabular}
  \caption{Sensitivity to \newRFsymbol bounds across multiple resources. For read–modify–write workloads, the end-to-end P99 latency stays within the additive bound \(\newRFsymbol_{\mathrm{total}} = \newRFsymbol_{\mathrm{buffer}} + \newRFsymbol_{\mathrm{cache}}\), confirming predictable composition of per-resource delay bounds.}
  \label{tab:multires-delta-sweep}
\end{table}

\subsubsection{Varying Workload Parameterization ($k$)}

We now vary the \peakdemandthreshold $k$, which captures the maximum number of clients that may reclaim their fair shares simultaneously in the system model. Intuitively, larger $k$ values allow for more concurrent ramp-up clients but require more conservative reservations, since the fixed refill bandwidth must be shared among more clients.

\autoref{tab:cache-k-sweep} reports results from the cache microbenchmark in \S\ref{ssec:eval-micro:cache} for $k \in \{1,\dots,6\}$ with a fixed \(\newRFsymbol=750\)~ms. As $k$ increases, the reservation fraction per client rises from 25\% at $k=1$ to 87.5\% at $k=6$, and throughput gradually decreases from 3.82~GB/s to 3.10~GB/s. This increase in reservation size reflects that, within a 750~ms refill window, only about 240~MB of data can be brought in from disk, and this limited refill budget must be divided among an ever larger set of ramp-up clients. Despite the reduced utilization at larger $k$, the observed P99 latency remains within the configured \newRFsymbol bound in all cases. These results show that \sys can be tuned to accommodate different assumptions about workload burstiness: smaller $k$ yields higher utilization under the assumption of few simultaneous ramp-ups, while larger $k$ protects against more concurrent ramp-ups by reserving more capacity per client.

\begin{table}[t]
  \centering
  \small
  \resizebox{\columnwidth}{!}{%
  \begin{tabular}{ccccc}
    \toprule
    \textbf{\newRFsymbol (ms)} & \textbf{$k$} & \textbf{Res (\%)} & \textbf{P99 Latency (ms)} & \textbf{Tput (GB/s)} \\
    \midrule
    750 & 1 &  25\%    & 729 & 3.82 \\
    750 & 2 & 62.5\%   & 739 & 3.53 \\
    750 & 3 & 75\%     & 741 & 3.41 \\
    750 & 4 & 81.25\%  & 726 & 3.29 \\
    750 & 5 & 85\%     & 734 & 3.18 \\
    750 & 6 & 87.5\%   & 732 & 3.10 \\
    \bottomrule
  \end{tabular}}
  \caption{Sensitivity to the system model parameter $k$. Increasing $k$ allows more ramp-up clients to reclaim their fair shares concurrently, which increases per-client reservations and reduces throughput, while P99 latency remains within the configured \newRFsymbol bound.}
  \label{tab:cache-k-sweep}
\end{table}

\subsubsection{Varying Workload Parameterization ($k$)}

\autoref{tab:cache-k-sweep} presents results from our cache microbenchmark across varying \peakdemandthreshold ($k = $1–6), which captures the maximum number of clients that may reclaim their fair share simultaneously. For instance, as $k$ increases, more ramping-up readers share the fixed read I/O bandwidth used to refill cache from disk, reducing each client's effective share and thus increasing the required cache reservations in the system. At $k=1$, only 25\% of each client’s cache must be reserved, but by $k=6$, this rises to 87.5\%. Given the \newRFsymbol=750ms refill deadline, only 240MB can be refilled—now divided among more clients. Despite this, all configurations meet the \newRFsymbol-bound on p99 latency, demonstrating that our policies preserve \newRFsymbol-fairness while adapting to different levels of workload burstiness.

%% file: 08-related.tex
\section{Related Work}
\label{sec:related-work}

\textbf{Fair scheduling.}
Our work is inspired by the long line of research from Competitive Equilibrium from Equal Incomes (CEEI)~\cite{ceei} in economics, to fair scheduling in networking, such as Start-time Fair Queuing (SFQ) in latency-sensitive settings~\cite{drfq-18}, and Generalized Processor Sharing (GPS)~\cite{demers1989analysis,drfq-24,drfq-11}. The SFQ algorithm aims to provide lower maximum delay for packets compared to its prior work, SCFQ~\cite{davin1990simulation,drfq-17} and WFQ~\cite{demers1989analysis,drfq-9}.
The GPS algorithms enable single- and multi-node networks to provide worst-case performance guarantees~\cite{drfq-24,drfq-11}. Dominant Resource Fair Queuing (DRFQ)~\cite{drfq} ensures the fair allocation of multiple exclusive resources for network packets. However, they all fairly schedule resources that exhibit negligible preemption delays. 

\noindent\textbf{Resource allocation in multi-tenant systems.}
A range of previous work addresses the fair allocation of resources in an OS, a DBMS, and a cluster~\cite{cake,mclock,ioflow,vdc,fair-ride,retro,pbox,pisces,sqlvm,drft}. For performance isolation across multiple tenants, variants of max-min fairness~\cite{waldspurger1994lottery} are popularly used to allocate a specific resource type, such as CPU~\cite{stoica1996proportional,waldspurger1994lottery}, memory~\cite{waldspurger1995stride,saraswat2007theory}, storage~\cite{axboe2004linux}, etc., or a fixed-size partition of resources~\cite{isard2009quincy}, and Dominant Resource Fairness (DRF)~\cite{drf} is widely-adopted for multiple resource types.
Retro~\cite{retro} implements various scheduling policies, including DRF, to achieve performance isolation of physical resources.
SQLVM~\cite{sqlvm} promises a fixed resource reservation to a tenant on SQL stores and Pisces~\cite{pisces} applies DRF to a cloud key-value storage service. In contrast to works that focus on mitigating the inference from shared physical resources, pBox~\cite{pbox} targets virtual resources (e.g., shared buffers, queues) and handles intra-application interference within the Linux kernel. 
Delay scheduling~\cite{zaharia2010delay} preserves data locality while scheduling jobs in a Hadoop cluster to improve performance.
However, these works do not address the large tail latency spikes in these systems or bound the delay in users receiving their fair share.
Other works that focus on providing performance isolation via latency SLAs~\cite{icbs}/SLOs~\cite{piql}, and predictable performance~\cite{tempo,bazaar}, often via admission control techniques~\cite{active-sla,acontrol-crdb}, complement our work.

%% file: 09-conclusion.tex
\section{Conclusion}

In this work, we address the problem of sharing resources with high preemption delays, a critical challenge in multi-tenant storage systems. We define the desired properties of \newRFfull algorithms that bound client delays to receive their fair share of resources and achieve high utilization. This work is distinct from prior fair sharing work due to the focus on high preemption delays and latency-sensitive settings.
We design and implement two \newRFfull algorithms in \sys, a prototype that extends RocksDB, and show that \sys reduces tail latency spikes end-to-end, meets clients' maximum delay targets, and reduces delays by up to 9$\times$ compared to traditional fair sharing on workloads from standard benchmarks.



\section{Acknowledgements}

This work is supported by Sky lab sponsors and affiliates, including Accenture, Amazon, AMD, Anyscale, Broadcom, Google, IBM, Intel, Intesa Sanpaolo, Lambda, Lightspeed, Mibura, NVIDIA, Samsung SDS, and SAP.
We thank Sunkanmi Adewumi for extending this work to distributed key-value stores, and Sriram Srivatsan for investigating the lack of performance isolation in PostgreSQL.
We also thank the members of the Sky and NetSys labs at UC Berkeley for their feedback, which significantly improved this work.